\documentclass[aps,prd,preprint,groupedaddress,amsmath,amssymb]{revtex4-2} 

\usepackage{hyperref}
\usepackage{graphicx}
\usepackage{dcolumn}
\usepackage{xspace}
\usepackage{caption}
\usepackage{subcaption}
\usepackage{bm}
\usepackage{hepunits}
\usepackage{siunitx}
\usepackage{hepnames}
\usepackage{booktabs}
\usepackage{multirow}

\newcommand*{\monster}{\textsc{MoNSTER}\xspace}
\newcommand*{\monsterempmode}{\textsc{MoNSTER} (Empirical Mode)\xspace}
\newcommand*{\monstermode}{\textsc{MoNSTER} (Mode)\xspace}
\newcommand*{\monstersample}{\textsc{MoNSTER} (Sample)\xspace}
\newcommand*{\modelfullname}{Mixture of Neutrino Solutions with Transformer Event Representation\xspace}

\newcommand*{\nuflows}{\ensuremath{\nu}-Flows\xspace}
\newcommand*{\nuflowssample}{\ensuremath{\nu}-Flows (Sample)\xspace}
\newcommand*{\nuflowsmode}{\ensuremath{\nu}-Flows (Empirical Mode)\xspace}

\newcommand*{\HPembeddim}{64\xspace}
\newcommand*{\HPencodernumlayers}{2\xspace}
\newcommand*{\HPenodernumheads}{8\xspace}

\newcommand*{\HPdropout}{0.01\xspace}
\newcommand*{\HPnumcomponents}{3\xspace}
\newcommand*{\HPlbfgsmaxiter}{100\xspace}
\newcommand*{\HPbatchsize}{128\xspace}
\newcommand*{\HPpeaklr}{0.001\xspace}
\newcommand*{\HPbetaone}{0.9\xspace}
\newcommand*{\HPbetatwo}{0.999\xspace}
\newcommand*{\HPweightdecay}{0.01\xspace}
\newcommand*{\HPwarmupfrac}{10\%\xspace}
\newcommand*{\HPmaxepochs}{100\xspace}
\newcommand*{\HPgradclip}{1\xspace}
\newcommand*{\deltap}{\ensuremath{\left| \Delta \vec{p} \right|}\xspace}
\newcommand*{\psixeight}{\ensuremath{P_{68}}\xspace}
\newcommand*{\pninefive}{\ensuremath{P_{95}}\xspace}

\begin{document}

\preprint{} 

\title{Mixture Density Networks for Neutrino Reconstruction at Hadron Colliders}

\author{Seungjin Yang}
\affiliation{Department of Physics, University of Seoul, Seoul, Republic of Korea}
\author{Jason S. H. Lee}
\email[]{jason.lee@yonsei.ac.kr}
\affiliation{Department of Physics, Yonsei University, Seoul, Republic of Korea}
\author{Junghwan Goh}
\affiliation{Department of Physics, Kyung Hee University, Seoul, Republic of Korea}

\date{\today}

\begin{abstract}
Neutrino momentum reconstruction at hadron colliders is intrinsically ambiguous because the longitudinal momentum is not directly observed.
We study this problem in semileptonic $t\bar{t}$ events using \monster{} (Mixture of Neutrino Solutions with Transformer Event Representation), a mixture density network that predicts a multivariate normal mixture for the conditional distribution of neutrino momentum from reconstructed event objects.
The model uses a Transformer-based event encoder and yields a sampling-free point estimate of the neutrino momentum from the closed-form density.
On the public benchmark introduced with \nuflows, \monster{} reduces the 68th percentile of the three-momentum residual by 5\% and the 95th percentile by 6\% relative to the empirical-mode \nuflows{} baseline, with per-variable root-mean-square errors smaller by 2 to 9\% at comparable bias, while running 2 to 8.5 times faster at inference, depending on the device and batch size.
These results indicate that mixture density networks are a competitive alternative to normalizing flows for neutrino reconstruction.
\end{abstract}


\maketitle

\section{Introduction}
\label{sec:intro}

At the Large Hadron Collider (LHC)~\cite{Evans:1129806}, proton-proton collisions produce complex multi-particle final states recorded as sets of reconstructed objects: jets, leptons, and missing transverse momentum ($\vec{p}_{T}^{\,\text{miss}}$).
Event reconstruction maps these detector-level observables to the underlying physical process, including the assignment of particle momenta and event topology.
Accurate reconstruction is a prerequisite for precision measurements and searches for new physics, since systematic uncertainties in object reconstruction propagate directly into the physics observables of interest.

Neutrinos escape the detector without interaction and leave no direct signal.
Their transverse momentum is inferred from the negative vector sum of all visible transverse momenta, which defines the missing transverse momentum $\vec{p}_{T}^{\,\text{miss}}$.
The longitudinal component is not directly observable and must be determined from additional kinematic constraints.
In semileptonic $t\bar{t}$ events, the on-shell $W$ boson mass constraint provides a quadratic equation relating the neutrino longitudinal momentum to the measured lepton kinematics and $\vec{p}_T^{\text{miss}}$.
Because the system is underconstrained by the unobserved longitudinal momentum, this quadratic equation yields a twofold ambiguity in the neutrino longitudinal momentum.

Deep learning has recently been applied to neutrino momentum reconstruction as a way to move beyond traditional single-valued kinematic solvers.
At first glance, casting neutrino momentum reconstruction as a regression (or point estimation) problem seems natural.
However, as pointed out in Ref.~\cite{Leigh:2022lpn}, the multimodal structure of the neutrino momentum, originating from missing information, causes deterministic regression models to collapse to a point between the two solutions, which is unphysical and leads to poor resolution.

To address this, Ref.~\cite{Leigh:2022lpn} introduced \nuflows{}, a conditional normalizing flow ~\cite{papamakarios2017masked,DBLP:conf/icml/RezendeM15} trained to model the full conditional distribution of the neutrino momentum given the event observables in semileptonic $t\bar{t}$ events.
Normalizing flows approximate the complex target distribution by transforming samples drawn from a simple base density through a sequence of invertible and differentiable mappings.
By framing the problem as conditional density estimation, \nuflows{} naturally accommodates the multimodal structure of the target distribution and allows diverse neutrino momentum candidates to be obtained by sampling.
Ref.~\cite{Raine:2023fko} extended this approach to multi-neutrino final states with $\nu^2$-Flows, demonstrating the generality of normalizing flow-based density estimation for this class of problems.

Besides methods that approximate complex distributions by transforming samples drawn from a simple distribution, there is also the mixture model approach, which approximates them by combining well-known simple distributions.
Mixture Density Networks (MDNs)~\cite{bishop1994mixture} are an early approach to neural conditional density estimation, based on the mixture model paradigm.
An MDN predicts the parameters of a mixture of parametric distributions as a function of the input, yielding a closed-form and differentiable density that can be evaluated without any sampling or Jacobian computation.
Unlike normalizing flows, which require each data point to pass through a chain of neural networks for density evaluation, an MDN exposes both the closed-form density and the mixture components directly after a single amortized network inference.
This enables sampling-free point estimators such as direct density optimization or selecting the most probable component, without the large-sample stochastic search that flow-based approaches require.

MDNs have been applied to a range of reconstruction and calibration tasks in high-energy physics experiments.
In the pixel detector at the ATLAS experiment~\cite{ATLAS:2008xda}, MDNs were studied as a replacement for the multi-stage system that used ten separate neural networks for cluster multiplicity classification and hit position regression~\cite{Khoda:2687968}.
This approach was adopted in Run 3, where three MDNs replaced the full Run 2 system, yielding narrower and more symmetric hit residuals and improving the performance of downstream Kalman filter-based track fitting~\cite{ATL-PHYS-PUB-2022-033}.
An MDN was also deployed as the adversary component in an adversarially trained jet tagger designed to decorrelate jet substructure discriminants from the reconstructed jet mass~\cite{ATL-PHYS-PUB-2018-014}.

For particle identification calibration at LHCb, an MDN was used to model the conditional distribution of particle identification classifier outputs as a function of particle kinematics and
detector occupancy, trained directly on weighted calibration data~\cite{Graziani:2021vai}.
The ATLAS collaboration applied an MDN for simultaneous energy and mass calibration of large-radius jets, introducing an asymmetric Gaussian loss to handle the non-Gaussian tails of the jet response~\cite{ATLAS:2023zca}.
On-chip deployment of MDNs has been explored for the High-Luminosity LHC, where compact architectures synthesized in 28 nm CMOS provide calibrated uncertainty estimates for hit positions and incident angles within nanosecond latency budgets~\cite{Das:2026mff}.

MDNs may be insufficient for the complex high-dimensional distributions in computer vision or language modeling, but the conditional distribution of neutrino momentum is low-dimensional and its multimodality is moderate in complexity.
We investigate whether an MDN-based approach can match or surpass the reconstruction performance of \nuflows{} while providing a sampling-free point estimate of the neutrino momentum.

We propose \monster{} (\modelfullname{}), a mixture density network for neutrino momentum reconstruction in semileptonic $t\bar{t}$ events at hadron colliders.
\monster{} uses a Transformer-based event encoder to process variable-length sets of reconstructed objects and maps the resulting event representation to the parameters of a multivariate normal mixture over the neutrino momentum.
Because the mixture density and its components are available explicitly, a single neutrino momentum estimate can be obtained either by direct gradient-based mode seeking or by selecting the most probable mixture component.
We compare \monster{} against \nuflows{}~\cite{Leigh:2022lpn} on the publicly available dataset from Ref.~\cite{zoch_2022_6782987}, and show that \monster{} achieves better resolution with comparable bias on this benchmark.

The remainder of this paper is organized as follows.
Section~\ref{sec:method} describes the mixture density network formalism, the model architecture, the mode-seeking procedure, the training configuration, the evaluation metrics, the dataset, and the baseline method.
Section~\ref{sec:result} presents the results.
Section~\ref{sec:conclusion} concludes.


\section{Method}
\label{sec:method}

\subsection{Mixture Density Networks}
\label{subsec:mdn}

A Mixture Density Network (MDN)~\cite{bishop1994mixture} is a neural network that models a conditional probability density $p(\mathbf{y}|\mathbf{x})$ as a mixture of parametric distributions, where the mixture parameters are predicted as functions of the input $\mathbf{x}$.
Rather than producing a single point estimate of the target $\mathbf{y}$, an MDN outputs a full probability distribution, enabling principled uncertainty quantification and the representation of multimodal or otherwise complex conditional densities.

In this work, the mixture components are taken to be multivariate normal distributions.
The conditional density of $D$ random variables $\mathbf{y} \in \mathbb{R}^D$ given input $\mathbf{x}$ is accordingly expressed as a $K$-component Multivariate Normal Mixture Model (MVNMM):
\begin{equation}
  p(\mathbf{y}|\mathbf{x}) = \sum_{k=1}^{K} \pi^{k}(\mathbf{x})\,
    \mathcal{N}\!\left(\mathbf{y} \mid \bm{\mu}^{k}(\mathbf{x}),\,
    \Sigma^{k}(\mathbf{x})\right),
    \label{eq:mvnmm}
\end{equation}
where $\pi^{k}(\mathbf{x})$ are the mixture weights, $\bm{\mu}^{k}(\mathbf{x}) \in \mathbb{R}^D$ are the component mean vectors, and $\bm{\Sigma}^{k}(\mathbf{x}) \in \mathbb{R}^{D \times D}$ are the component covariance matrices, all predicted by the network as functions of $\mathbf{x}$.
The mixture weights satisfy $\pi^{k} > 0$ and $\sum_{k=1}^{K} \pi^{k}(\mathbf{x}) = 1$ for all $\mathbf{x}$, and each $\Sigma^{k}(\mathbf{x})$ is constrained to be symmetric positive definite, ensuring that each component defines a valid multivariate normal distribution.

The MVNMM in Eq.~\eqref{eq:mvnmm} possesses several properties that make it well suited for neutrino momentum reconstruction.
First, the mixture density has a closed-form expression, making the negative log-likelihood (NLL) tractable and differentiable, so the network can be trained end-to-end by direct minimization of the NLL.
Second, by summing over $K$ components, the model can represent multimodal conditional densities, which arise naturally in semileptonic $t\bar{t}$ events due to the discrete ambiguity in kinematic solutions for the neutrino longitudinal momentum.
Third, because the component distributions are multivariate normals, the full covariance structure among the neutrino momentum components is captured by each $\Sigma^{k}$, rather than assuming independence across dimensions.
Finally, the closed-form density permits direct mode seeking via gradient-based optimization on $p(\mathbf{y}|\mathbf{x})$.


\subsection{Model Architecture}
\label{subsec:model}

\begin{figure}
  \centering
  \begin{subfigure}[b]{0.3\textwidth}
    \centering
    \includegraphics[width=\textwidth]{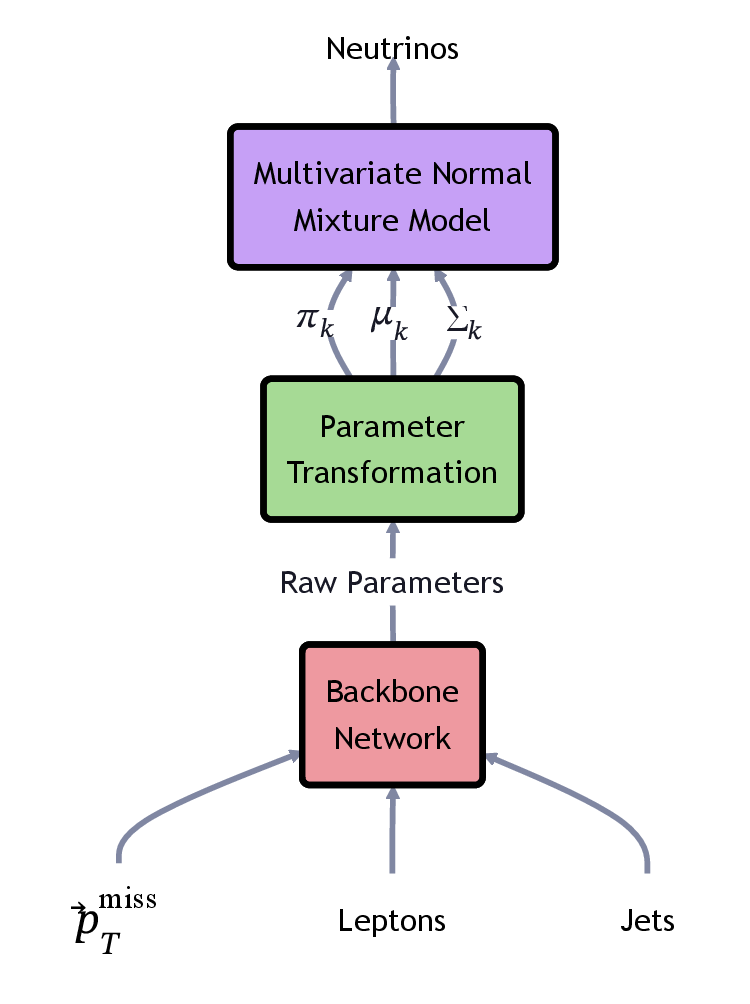}
    \caption{Architecture overview.}
    \label{fig:model-overview}
  \end{subfigure}
  \hfill
  \begin{subfigure}[b]{0.3\textwidth}
    \centering
    \includegraphics[width=\textwidth]{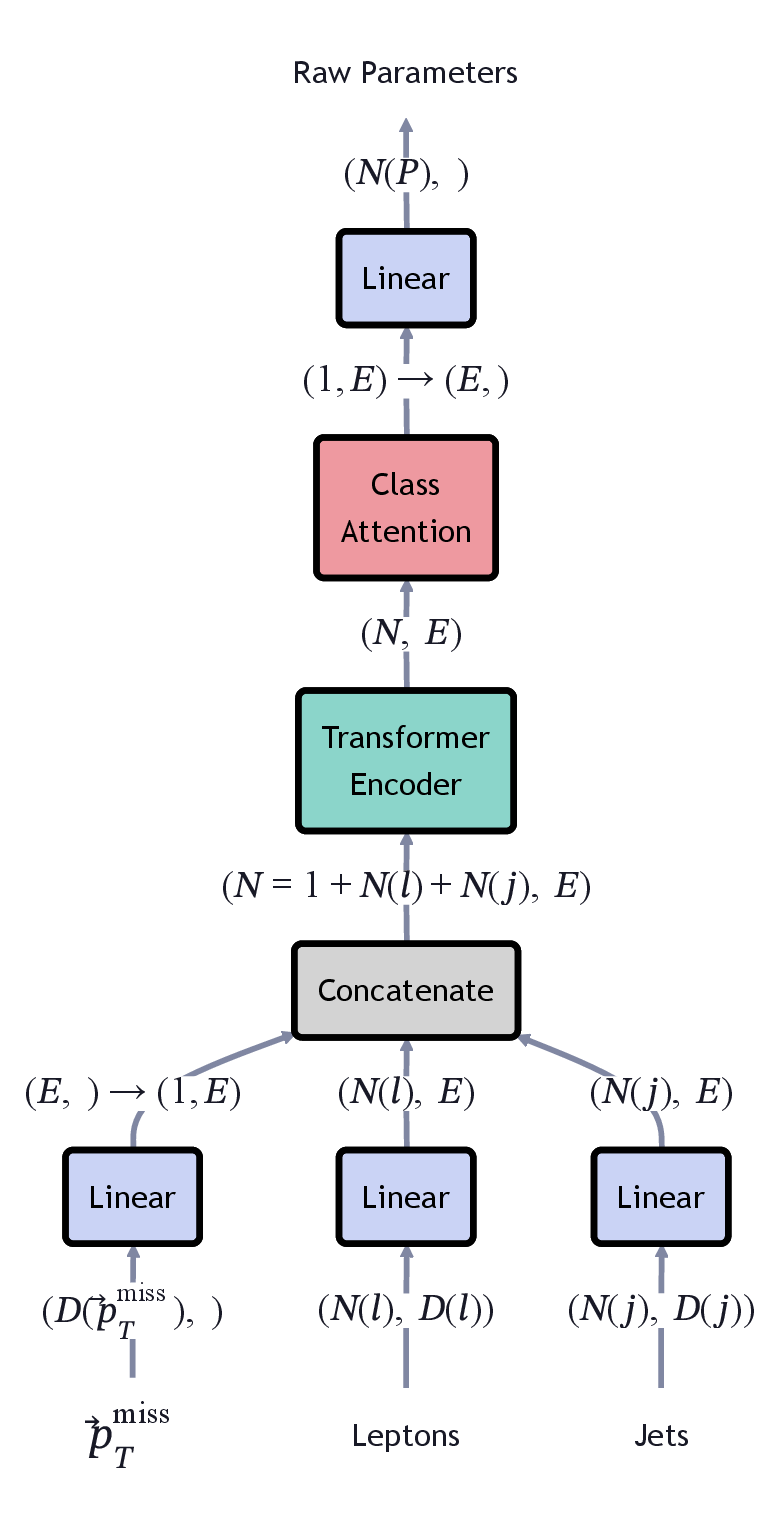}
    \caption{Encoder and class attention.}
    \label{fig:model-net}
  \end{subfigure}
  \hfill
  \begin{subfigure}[b]{0.3\textwidth}
    \centering
    \includegraphics[width=\textwidth]{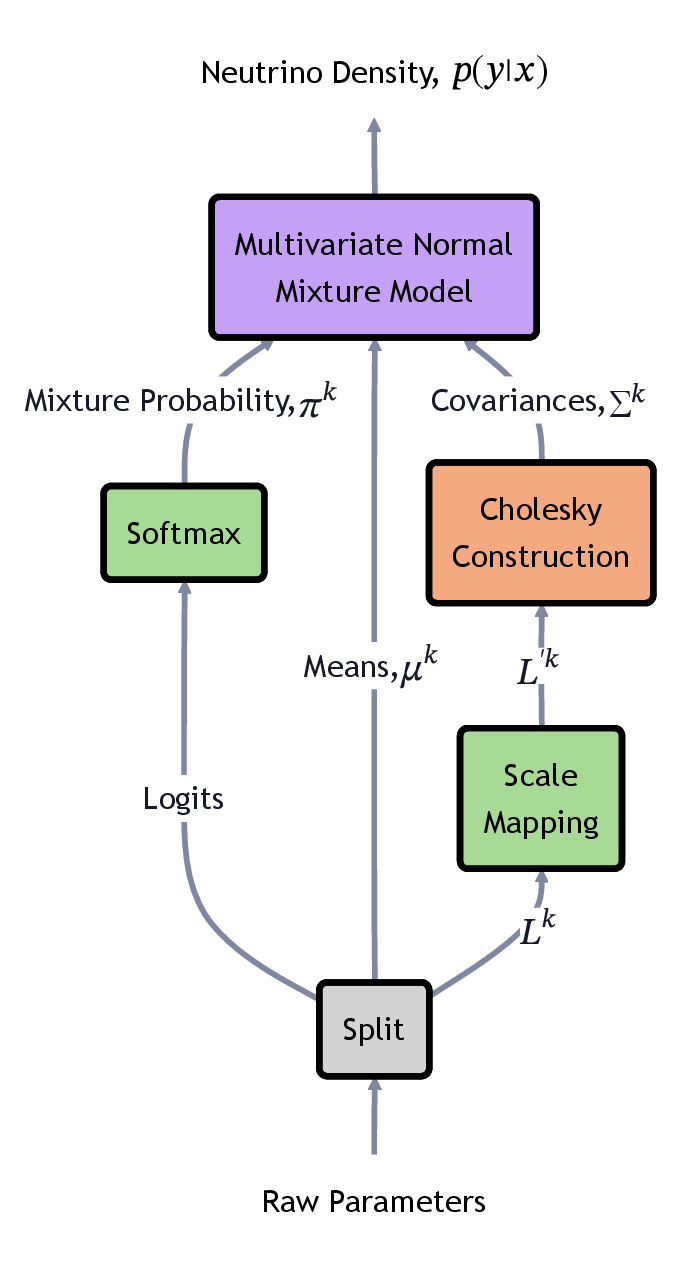}
    \caption{MVNMM parameter construction.}
    \label{fig:model-mvnmm}
  \end{subfigure}
  \caption{
    Architecture of the \monster{} network.
    (a) The full pipeline from reconstructed event objects to the predicted neutrino momentum distribution.
    (b) The Transformer encoder and class-attention module that produce a fixed-size latent event vector.
    (c) Transformation of the raw linear-layer output into the mixture weights, mean vectors, and covariance matrices of the MVNMM.
  }
  \label{fig:model}
\end{figure}

Figure~\ref{fig:model} illustrates the overall architecture of \monster.
Conceptually, \monster{} comprises two stages: a regression network that predicts the raw parameters of a mixture distribution over the neutrino momentum, and the construction of the MVNMM from these raw parameters via deterministic transformations that ensure the validity of the mixture parameters.
The target variables are defined as $\mathbf{y} = \left(p_{x}(\nu),\, p_{y}(\nu),\, \eta(\nu)\right)$, following Ref.~\cite{Leigh:2022lpn}.

\begin{table}
  \caption{Per-object input features used by the regression network.}
  \label{tab:input}
  \centering
  \small
  \begin{tabular*}{\columnwidth}{@{\extracolsep{\fill}}lll@{}}
    \toprule
    Input object & Features & Description \\
    \midrule
    \shortstack[l]{$\vec{p}_{T}^{\,\text{miss}}$} &
    \shortstack[l]{$p_{x}^{\text{miss}},\, p_{y}^{\text{miss}}$} &
    \shortstack[l]{Missing transverse\\momentum components} \\
    \midrule
    \shortstack[l]{Charged Lepton ($l$)} &
    \shortstack[l]{$p_{x}(l),\, p_{y}(l),\, \eta(l), \log E(l),\, I_{\mu}$} &
    \shortstack[l]{Kinematics and binary\\muon flavor indicator} \\
    \midrule
    \shortstack[l]{Jet ($j$)} &
    \shortstack[l]{$p_{x}(j),\, p_{y}(j),\, \eta(j), \log E(j),\, I_{b-\text{tag}}$} &
    \shortstack[l]{Kinematics and binary\\$b$-tag information} \\
    \bottomrule
  \end{tabular*}
\end{table}

The regression network takes as input a variable-length set of reconstructed objects in each event and outputs the raw parameters from which the MVNMM is built.
The input consists of the charged lepton, the reconstructed jets, and the missing transverse momentum vector, with each object represented by the features listed in Tab.~\ref{tab:input}.
\monster{} uses every jet stored for each event, up to ten.
This input definition follows \nuflows{}~\cite{Leigh:2022lpn} to enable a direct comparison, except that \monster{} does not use the jet and $b$-tagged jet multiplicities.
Continuous kinematic features are standardized to zero mean and unit variance, while binary features remain unchanged.

\begin{figure}
  \centering
  \begin{subfigure}[b]{0.32\textwidth}
    \centering
    \includegraphics[height=0.5\textheight]{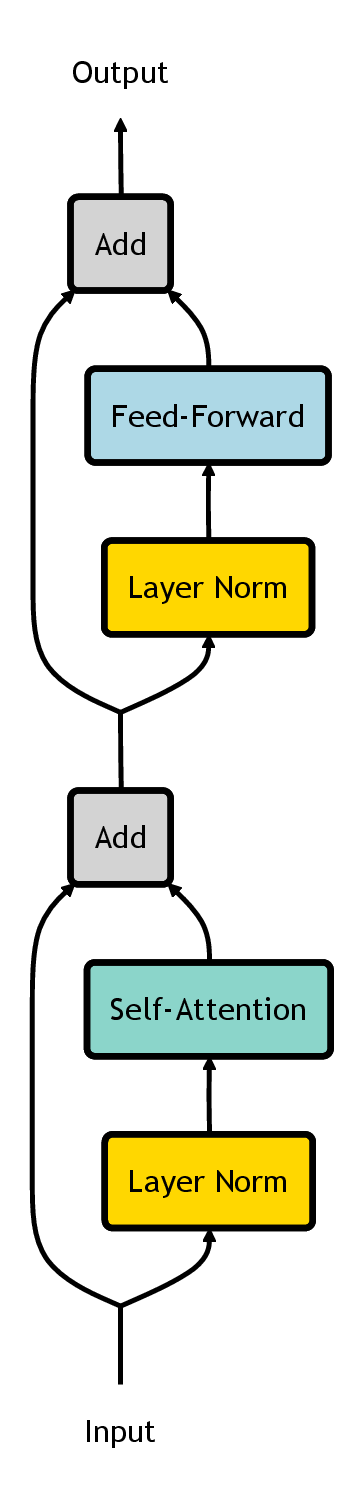}
    \caption{}
    \label{fig:model-encoder-layer}
  \end{subfigure}
  \hfill
  \begin{subfigure}[b]{0.32\textwidth}
    \centering
    \includegraphics[height=0.5\textheight]{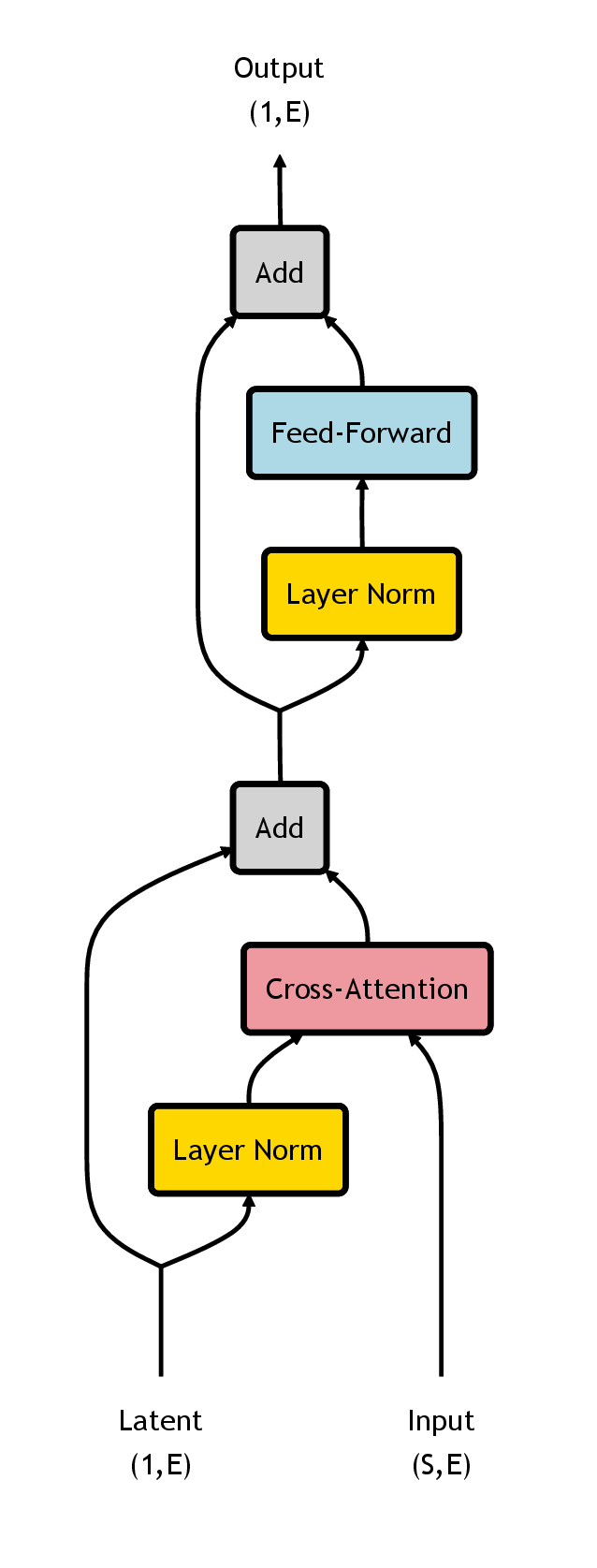}
    \caption{}
    \label{fig:model-cls-attn}
  \end{subfigure}
  \hfill
  \begin{subfigure}[b]{0.32\textwidth}
    \centering
    \includegraphics[height=0.35\textheight]{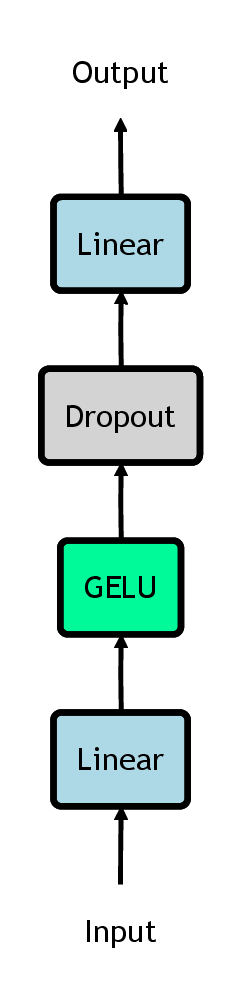}
    \caption{}
    \label{fig:model-ffn}
  \end{subfigure}
  \caption{
    Components of the regression network.
    (a) The pre-layer-normalized Transformer encoder layer used to contextualize the per-object embeddings through self-attention and a position-wise feed-forward network (FFN).
    (b) The class-attention module, where a single learnable latent query attends to the encoder outputs through cross-attention to produce an event-level representation.
    (c) The FFN block used in both the encoder and class-attention modules.
  }
  \label{fig:model-component}
\end{figure}

Figure~\ref{fig:model-net} illustrates the data flow of the regression network.
The network consists of input projection layers, a Transformer encoder, a class-attention module, and a final linear layer.
First, each object is projected into an $E$-dimensional embedding space via a dedicated linear layer.
The resulting embeddings are then stacked into a single sequence.

The embedding array is passed through a stack of $N$ identical encoder layers, each comprising a multi-head self-attention sub-layer followed by a position-wise feed-forward network (FFN) sub-layer, as shown in Fig.~\ref{fig:model-encoder-layer}.
We adopt a pre-layer normalization architecture, which places layer normalizations before each sub-layer rather than after, stabilizing gradient flow and improving convergence~\cite{DBLP:conf/icml/XiongYHZZXZLWL20}.
The self-attention sub-layer allows every object in the event to attend to every other object, producing a contextualized representation of the full event.

The FFN sub-layer, illustrated in Fig.~\ref{fig:model-ffn}, transforms each position independently.
First, a linear projection expands the embedding dimension by a factor of four, after which a Gaussian Error Linear Unit (GELU) activation~\cite{DBLP:journals/corr/HendrycksG16} is applied.
Dropout~\cite{DBLP:journals/jmlr/SrivastavaHKSS14} with rate \HPdropout is applied for regularization.
A second linear layer returns the representation to the original embedding dimension.

The Transformer encoder preserves the input array length and returns one contextualized embedding for each input object.
The regression head, however, must produce a fixed-size vector of MVNMM parameters.
We therefore aggregate the variable-length encoder output with a class-attention mechanism inspired by the class token in vision Transformers~\cite{DBLP:journals/corr/abs-2010-11929}.
The module uses a single learnable query vector, while the full sequence of encoder outputs provides the keys and values.
In practice, it is implemented as a Transformer-style decoder block with pre-layer normalization, a cross-attention sub-layer, and an FFN, as shown in Fig.~\ref{fig:model-cls-attn}.
The self-attention sub-layer of the standard decoder is omitted because the query consists of only one vector.
The resulting output is a latent event representation $\bm{z} \in \mathbb{R}^{E}$.

The latent representation $\bm{z}$ is mapped by a final linear layer to a raw parameter vector for the MVNMM.
Its dimension is $K \times \left(1 + D + D(D+1)/2\right)$, since each mixture component contributes one weight logit, $D$ mean parameters, and $D(D+1)/2$ entries for a lower-triangular matrix.
In our case, $D=3$.
The output of the final linear head is partitioned into three groups and transformed into the MVNMM parameters as follows, as illustrated in Fig.~\ref{fig:model-mvnmm}.
First, $K$ logits are passed through a softmax to yield the mixture weights $\{\pi^{k}\}_{k=1}^{K}$, satisfying $\pi^{k} > 0$ and $\sum_k \pi_k = 1$.
Second, the next $K \times D$ values are taken directly as the component means $\{\bm{\mu}^{k}\}_{k=1}^{K}$.
Third, the remaining $K \times D(D+1)/2$ values are arranged into $K$ lower-triangular matrices $\{L^{k}\}_{k=1}^K$.
The diagonal entries of each $L^k$ are mapped to strictly positive values via a biased softplus function~\cite{NIPS2000_44968aec}, while the off-diagonal entries are left unconstrained:
\begin{equation}
  L'^{k}_{ij}
  =
  \begin{cases}
    \mathrm{softplus}\!\left(L^{k}_{ij} + \mathrm{softplus}^{-1}(1 - \epsilon) \right) + \epsilon, & i = j,\\
    L^{k}_{ij}, & i \neq j,
  \end{cases}
  \label{eq:biased-softplus}
\end{equation}
where $\mathrm{softplus}(x) = \log(1 + e^x)$ and $\epsilon = 0.01$ is a small positive constant that prevents the diagonal entries from vanishing.
The covariance matrix of the $k$th component is then assembled from its Cholesky factor as $\Sigma^{k} = L'^{k}(L'^{k})^{\top}$. As $L'^{k}$ has strictly positive diagonal entries, every $\Sigma^{k}$ is symmetric positive definite by construction.

\subsection{Mode Seeking}
\label{subsec:mode-seeking}

To retrieve a single point estimate of the neutrino momentum from the predicted MVNMM, we use the mode of the conditional density $p(\mathbf{y}|\mathbf{x})$,
\begin{equation}
  \hat{\mathbf{y}}_{\mathrm{mode}} = \arg\max_{\mathbf{y}} p(\mathbf{y} \mid \mathbf{x}).
  \label{eq:mode}
\end{equation}
Because $p(\mathbf{y} \mid \mathbf{x})$ is available in closed form, this mode seeking can be carried out directly in the three-dimensional neutrino momentum space.

The objective is non-convex, so we use a multi-start strategy based on the component means of the mixture.
For each event, we initialize the optimizer from the $K$ predicted means $\{\bm{\mu}^{k}\}_{k=1}^{K}$ and run $K$ parallel instances of the Limited-memory Broyden--Fletcher--Goldfarb--Shanno (L-BFGS) algorithm~\cite{liu1989limited} to minimize the NLL.
This choice is well motivated because the high-density regions of a normal mixture are typically located near the component centers.
We use the PyTorch implementation with its default learning rate of 1 and strong Wolfe line search~\cite{wolfe1969convergence,wolfe1971convergence,wright1999numerical}. 
Among the $K$ optimization runs, we retain the converged candidate with the smallest NLL as the final mode estimate.
This procedure provides an approximation to the conditional mode.
We refer to this approach as the \textit{L-BFGS estimator}.

Alternatively, one could use the mean of the component with the largest mixture weight as a point estimate~\cite{bishop1994mixture}, which avoids the overhead of iterative optimization.
We denote this approach as the \textit{Most Probable Component estimator}.
When the mixture weights are well separated, this approach can yield a good approximation to the mode, but it may fail when the mixture density exhibits a complex and multimodal landscape.

Because the choice of mode-seeking method depends on the complexity of the mixture density, we defer selecting the mode-seeking strategy until after determining the number of mixture components, $K$, which is discussed in Sec.~\ref{subsec:model-selection}.
The two estimators are compared in Sec.~\ref{subsec:mode-seeking-comparison}, and the estimator adopted for the final results is selected there.
For the model-selection scan of Sec.~\ref{subsec:model-selection} we fix a single estimator, using L-BFGS with a maximum of \HPlbfgsmaxiter iterations per initialization, so that models with different $K$ are compared on equal footing.

\subsection{Training Configuration}
\label{subsec:training}

We train \monster{} by minimizing the NLL of the target neutrino momentum under the predicted MVNMM.
Each optimization step uses a batch of \HPbatchsize events.
Parameter updates are performed with the AdamW optimizer~\cite{DBLP:journals/corr/abs-1711-05101} with exponential decay rates $\beta_1 = \HPbetaone$ and $\beta_2 = \HPbetatwo$ for the first- and second-moment estimates.
Weight decay with coefficient \HPweightdecay is applied only to the weight matrices of the linear and attention layers, and not to bias terms or layer-normalization parameters.
To improve numerical stability, the gradient norm is clipped to \HPgradclip.

We use a learning-rate schedule with a linear warmup phase~\cite{DBLP:journals/corr/GoyalDGNWKTJH17} followed by cosine annealing~\cite{DBLP:journals/corr/LoshchilovH16a}.
The learning rate increases linearly from zero to a peak value of \HPpeaklr over the first \HPwarmupfrac of the total training epochs, and then decreases smoothly to zero by the end of training.
This schedule improves optimization stability in the early stage of training while avoiding abrupt changes in the later stage.
Training is run for \HPmaxepochs epochs.

\subsection{Evaluation Metrics}
\label{subsec:metrics}

We evaluate reconstruction performance through the norm of the three-momentum residual vector,
\begin{equation}
    \deltap = \left\| \hat{\vec{p}}(\nu) - \vec{p}^{\,\mathrm{truth}}(\nu) \right\|_{2}
    \label{eq:dp}
\end{equation}
where $\hat{\vec{p}}(\nu)$ denotes the reconstructed neutrino three-momentum and $\vec{p}^{\,\mathrm{truth}}(\nu)$ the generator-level value.
By construction, $\deltap$ is non-negative and vanishes only for a perfect reconstruction, so smaller values indicate better agreement with the truth.
Combining all three momentum components into a single scalar yields a measure of the overall reconstruction quality and avoids ambiguities in how per-component residuals should be weighted against one another.

To summarize the distribution of $\deltap$, we report its $k$-th percentile, $P_{k}$.
We use \psixeight{} as the primary summary metric for comparing reconstruction methods.
For model selection, we evaluate model checkpoints on the validation set using \psixeight{} and retain the checkpoint with the lowest value.
This choice reflects our focus on accurate point reconstruction, whereas NLL measures the quality of the full conditional density in the normalized space of the target variables.

\subsection{Dataset}
\label{subsec:dataset}

We use the publicly available dataset~\cite{zoch_2022_6782987} of simulated semileptonic $t\bar{t}$ events from Ref.~\cite{Leigh:2022lpn}.
Proton--proton collisions are simulated at a center-of-mass energy of 13 TeV, with the top quark mass set to $\SI{173}{\giga\electronvolt}$ and the $W$ boson mass set to $\SI{80.38}{\giga\electronvolt}$.
Hard interactions are modeled at leading order using MadGraph5~\cite{Alwall:2014hca}, with top quark and $W$ boson decays handled by \textsc{MadSpin}~\cite{Artoisenet:2012st}.
Parton showering and hadronization are performed with \textsc{Pythia8}~\cite{Sjostrand:2007gs}.
The detector response is simulated with \textsc{Delphes}~\cite{deFavereau:2013fsa} using the default ATLAS detector configuration.

Jets are reconstructed from energy-flow objects using the anti-$k_{t}$ algorithm~\cite{Cacciari:2008gp} with $R=0.4$.
Jets are $b$-tagged with an efficiency parameterized as a function of jet $p_T$, at a working point with 70\% efficiency for true $b$ jets. Events are required to have at least four jets with $p_{T} > 25$ GeV and $|\eta| < 2.5$, and exactly one reconstructed electron or muon with $p_{T} > 15$ GeV and $|\eta| < 2.5$.
At most ten jets are stored per event, those with the highest $p_T$, ordered by decreasing $p_T$.

The dataset contains 585,600 training events and 94,293 test events.
Of the training set, 20\% is reserved for validation and model selection, and the remaining 80\% is used for model training.
The test set is used only for the final evaluation of the trained model.


\subsection{Reference Method}
\label{subsec:baseline}
To benchmark \monster{}, we compare against \nuflows{}, a conditional normalizing-flow method for neutrino regression introduced in Ref.~\cite{Leigh:2022lpn}.
The conditioning inputs comprise the missing transverse momentum components, the charged lepton kinematics and flavor indicator, the kinematic and $b$-tag information of all stored jets, and the jet and $b$-tagged jet multiplicities.

Architecturally, $\nu$-Flows consists of a feed-forward conditioning network followed by a conditional invertible neural network (cINN)~\cite{DBLP:journals/corr/abs-1907-02392}.
In the conditioning network, the jet collection is processed with an attention-weighted DeepSet~\cite{DBLP:journals/corr/ZaheerKRPSS17}, which preserves permutation invariance while accommodating variable jet multiplicity.
This jet-level summary is combined with the non-jet observables through an embedding network to produce the context tensor supplied to the flow layers.

The cINN itself is built from seven alternating conditional coupling layers using rational-quadratic spline transformations~\cite{DBLP:conf/nips/DurkanB0P19}, together with LU-decomposed linear transformations~\cite{NEURIPS2018_d139db6a} between flow blocks.
In the first coupling layer, the neutrino variables are partitioned such that the transverse components are assigned to one subset and the longitudinal coordinate to the other, and this masking is alternated in subsequent layers.
The dense subnetworks used in the DeepSet, the embedding network, and the spline networks employ two hidden layers with 64 nodes each, LeakyReLU activations~\cite{DBLP:journals/corr/XuWCL15}, layer normalization~\cite{DBLP:journals/corr/BaKH16}, and residual connections~\cite{DBLP:journals/corr/HeZRS15,DBLP:journals/corr/HeZR016}.

In this work, we use the official public implementation of \nuflows{}~\cite{Leigh_neutrino_flows} and train it on the same dataset as \monster{} for a fair comparison.
We train \nuflows{} ten times with different random seeds and use the model with the lowest validation \psixeight{}.
For point-estimate evaluation, we use the empirical-mode estimator, denoted as \nuflowsmode{}, obtained by drawing 256 conditional samples per event and retaining the sample with the highest conditional density, following Ref.~\cite{Leigh:2022lpn}.
We also use a single sample as a point estimate, denoted as \nuflowssample{}.

To aid interpretation, we also compute the kinematic solutions implied by the on-shell $W$ boson mass constraint, which is widely used in traditional neutrino reconstruction methods~\cite{ATLAS:2015pfy,CMS:2018quc,Kvita:2018trd,ATLAS:2018fwq,ATLAS:2019guf,ATLAS:2019hxz,Grossi:2020orx,CMS:2021vhb,ATLAS:2022waa,Leigh:2022lpn}.
\begin{equation}
    \label{eq:kinematic-solution}
    p_{z}(\nu) = \frac{-b \pm \sqrt{b^2-4ac}}{2a},
\end{equation}
where
\begin{flalign*}
  a &= p_{z}^{2}\left(l\right) - E^{2}\left(l\right), \\
  b &= \alpha p_{z}\left(l\right), \\
  c &= \frac{\alpha^2}{4}-E^{2}\left(l\right)p_{\mathrm{T}}^{2}(\nu), \\
  \alpha &= m^{2}\left(W\right) - m^{2}\left(l\right) + 2\left(p_{x}\left(l\right) p_{x}\left(\nu\right) + p_{y}\left(l\right) p_{y}\left(\nu\right) \right).
\end{flalign*}
Following the simulation setup, the $W$ boson mass is fixed to $\SI{80.38}{\giga\electronvolt}$.
This yields two real solutions when the discriminant is positive and a single degenerate solution when it vanishes.
In the two-solution case, we choose the solution with the smaller $|p_{z}(\nu)|$, which is a common heuristic in traditional reconstruction methods.
When the discriminant is negative, the two roots are complex conjugates sharing the real part $-b/2a$, and we take that value as the single solution.

\section{Results}
\label{sec:result}

\subsection{Model Selection}
\label{subsec:model-selection}

\begin{table}
  \caption{%
    The 68th and 95th percentiles (\psixeight{} and \pninefive{}) of the neutrino momentum residual are shown for different numbers of mixture components (K) in the multivariate normal mixture model.
    Each model was trained for 100 epochs.
    The reported values are the mean and standard deviation computed over ten independent runs with different random seeds, evaluated on the validation set.
  }
  \label{tab:hpo}
  \begin{center}
\begin{tabular}{crrrrrrr}
\toprule
K & 2 & 3 & 4 & 5 & 6 & 7 & 8 \\
\midrule
  $P_{68}$ [GeV] & 61.7$\pm$0.4 & 61.0$\pm$0.3 & 60.8$\pm$0.4 & 60.5$\pm$0.7 & 60.5$\pm$0.8 & 60.8$\pm$0.3 & 60.6$\pm$0.4 \\
  $P_{95}$ [GeV] & 233.6$\pm$1.3 & 233.4$\pm$1.7 & 234.8$\pm$1.0 & 235.7$\pm$1.5 & 234.9$\pm$2.2 & 236.2$\pm$1.6 & 236.0$\pm$1.8 \\
\bottomrule
\end{tabular}
  \end{center}
\end{table}

We perform hyperparameter optimization over the number of mixture components $K$ in the MVNMM, which controls the capacity of the model to represent multimodality in the conditional density.
Since small networks already perform well, we fix the regression architecture during the scan to isolate the effect of $K$.
The network uses an embedding dimension $E=\HPembeddim$.
The Transformer encoder has $\HPencodernumlayers$ layers with $\HPenodernumheads$ attention heads, and the class-attention module has one layer with a single attention head.
For this dedicated scan, we train models with $K=2,\ldots,8$ for 100 epochs with ten different random seeds for each $K$.
The average training time is about 2 hours on a single NVIDIA GeForce RTX 4090 GPU.

Table~\ref{tab:hpo} shows the results of this scan.
Models with $K \geq 3$ give similar validation performance, with the validation \psixeight{} between 60.5 and 61.0 GeV.
The lowest mean \psixeight{} is obtained for $K=5$ and $K=6$, but the differences among $K \geq 3$ are comparable to the seed-to-seed variation.
The \pninefive{} values show no corresponding improvement at larger $K$, and the smallest mean \pninefive{} is obtained for $K=3$.

The two-component model is worse in \psixeight{}, despite the twofold ambiguity in the kinematic solutions for the neutrino pseudorapidity.
This indicates that the twofold kinematic ambiguity does not translate into an exactly two-component conditional density at reconstruction level.
A third component likely provides additional flexibility to model smearing and branch overlap, while further increasing $K$ yields no statistically meaningful gain in validation performance.
We choose $K=\HPnumcomponents{}$ for the final model because it is statistically compatible with the lowest \psixeight{}, shows no degradation in \pninefive{} relative to any other value of $K$, and keeps the mixture compact.
The resulting network has about 153,000 trainable parameters.

\subsection{Mode Seeking Comparison}
\label{subsec:mode-seeking-comparison}

\begin{figure}
  \begin{center}
    \includegraphics[width=0.32\textwidth]{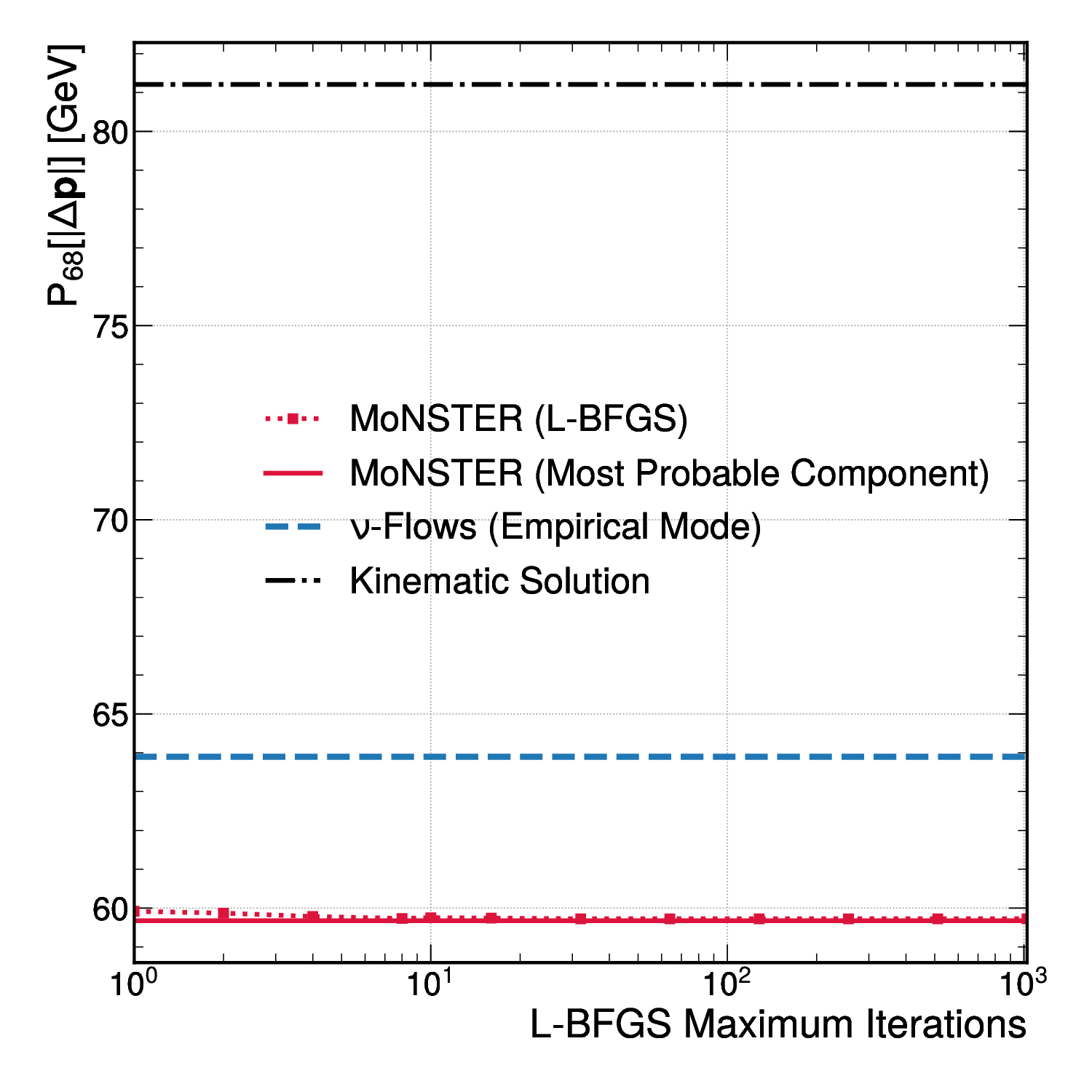}
    \includegraphics[width=0.32\textwidth]{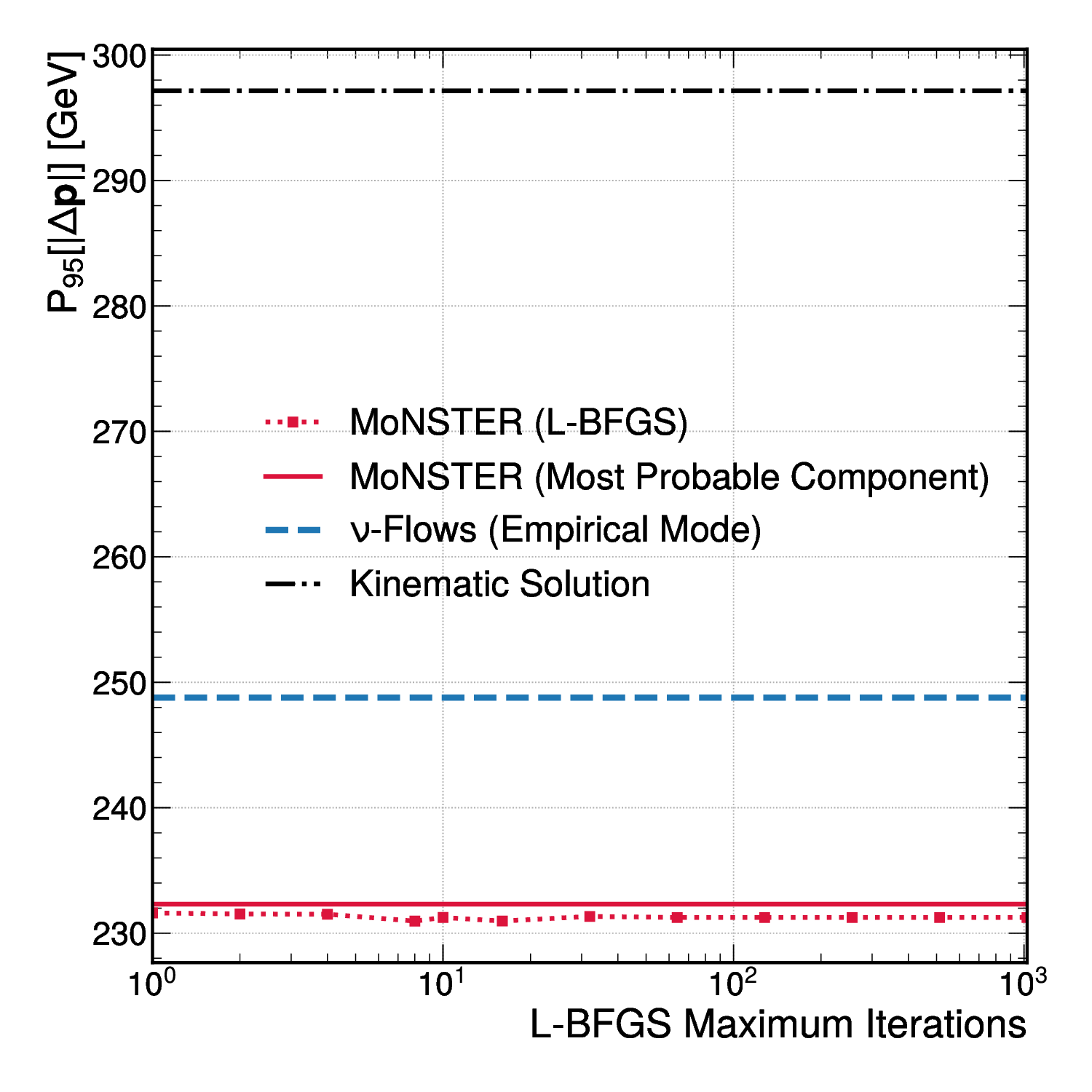}
    \includegraphics[width=0.32\textwidth]{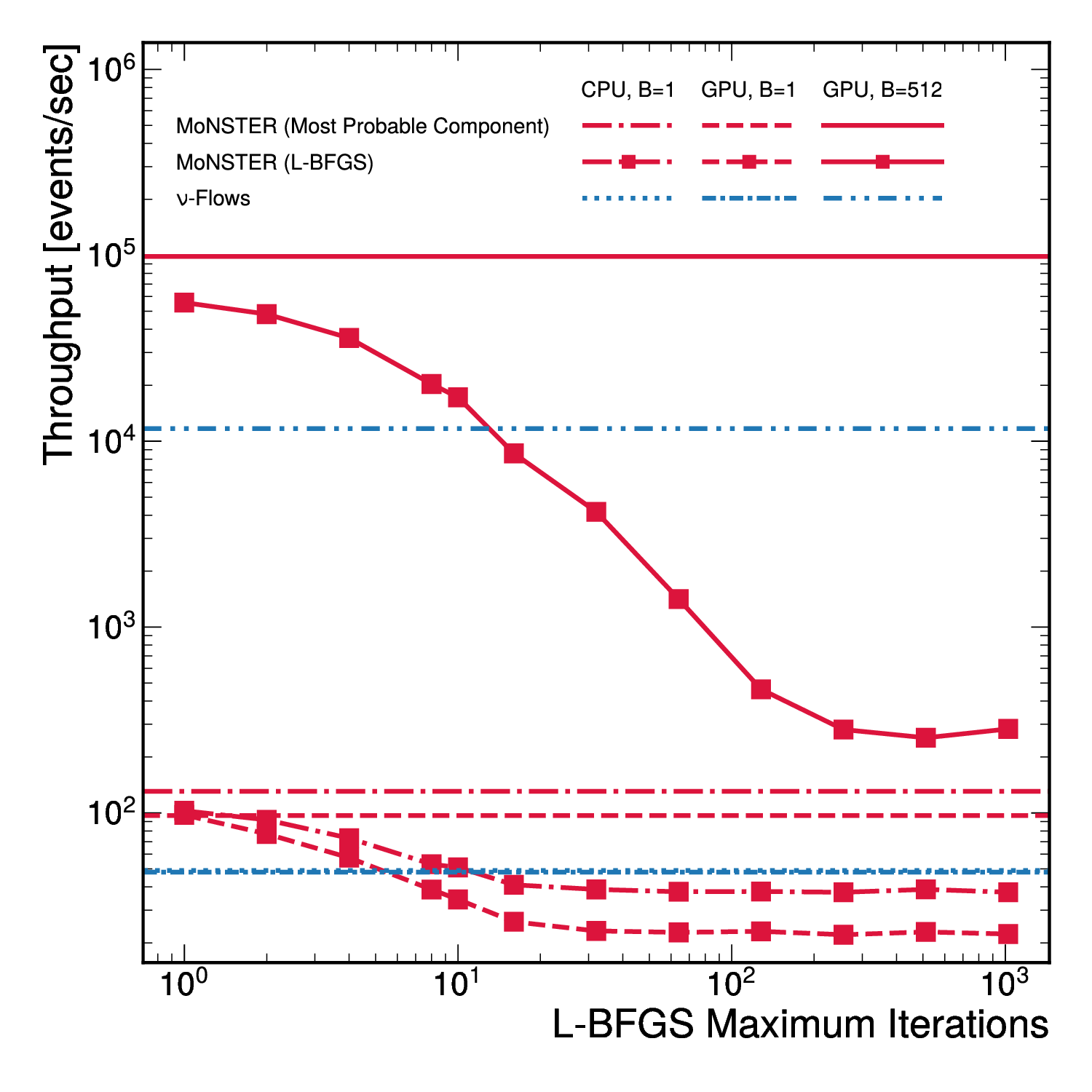}
  \end{center}
  \caption{%
    Dependence of reconstruction performance and inference throughput on the maximum number of L-BFGS iterations used for mode seeking.
    The left and center panels show, respectively, the 68th and 95th percentiles of the neutrino three-momentum residual $\deltap$ on the validation set.
    Both \monster{} estimators use a single checkpoint: the one with the lowest validation \psixeight{} evaluated with L-BFGS among the $K=\HPnumcomponents{}$ runs of the dedicated scan.
    The \nuflows{} reference uses the single model selected by validation \psixeight{} as described in Sec.~\ref{subsec:baseline}.
    The red squares denote \monster{} modes obtained with L-BFGS, while the horizontal lines show the most probable mixture component (red), \nuflowsmode{} (blue), and the kinematic solution (black) for comparison.
    The right panel shows the throughput of the Most Probable Component and L-BFGS estimators of \monster{}, together with \nuflowsmode{}, for CPU and GPU inference at the batch sizes ($B$) indicated in the legend.
  }
  \label{fig:mode-seeking-comparison}
\end{figure}

Figure~\ref{fig:mode-seeking-comparison} compares the Most Probable Component estimator with the L-BFGS estimator, both evaluated on the validation set.
The Most Probable Component estimator returns the mean of the mixture component with the largest predicted weight and requires no numerical optimization.
The L-BFGS estimator instead performs the multi-start optimization described in Sec.~\ref{subsec:mode-seeking}, which is initialized from the $K$ component means and retains the candidate with the smallest NLL.
It is evaluated here as a function of the maximum number of iterations allowed per initialization.

Increasing the maximum number of L-BFGS iterations has little effect on reconstruction performance.
Both \psixeight{} and \pninefive{} vary by well under 1\% across the full range of the scan, and \pninefive{} does not decrease monotonically with the iteration limit.
The optimization therefore reaches essentially the same modes after very few iterations, and further iterations do not translate into improved reconstruction.

The Most Probable Component estimator remains competitive with the L-BFGS estimator across the entire scan.
It attains a \psixeight{} at least as small as any value reached by L-BFGS, at the cost of a marginally wider upper tail as measured by \pninefive{}.
The two estimators are therefore comparable in accuracy for practical purposes, while the Most Probable Component estimator dispenses with iterative optimization entirely.
Both \monster{} estimators also give smaller percentile values than \nuflowsmode{} and the kinematic solution shown in the figure.

The throughput measurements further favor the Most Probable Component estimator.
Throughput was measured on a system separate from the one used for training, with Intel Xeon Gold 5218 CPUs and a single NVIDIA GeForce RTX 5080 GPU. All methods were timed on this same system, so only relative throughputs are compared.
For GPU inference with a batch size of 512, it is 1.8 times faster than L-BFGS with an iteration limit of one and 8.5 times faster than \nuflows{}.
For CPU inference with a batch size of one, it is 1.3 times faster than L-BFGS with an iteration limit of one and 2.7 times faster than \nuflows{}.
At a batch size of one on the GPU, the two \monster{} estimators have comparable throughput, and the Most Probable Component estimator is approximately twice as fast as \nuflows{}.

Its computational advantage over L-BFGS increases with the iteration limit because it requires no numerical optimization.
At a batch size of one, however, each event entails too little numerical work to amortize CUDA dispatch and synchronization overhead, so CPU inference is comparable to, or even faster than, GPU inference.

We therefore use the Most Probable Component estimator for neutrino momentum reconstruction.
It matches the best \psixeight{} obtained with L-BFGS, retains competitive \pninefive{} performance, and avoids the computational cost of iterative optimization.
In the remainder of this paper, \monstermode{} denotes \monster{} evaluated with this estimator.


\subsection{
Performance Across Training Seeds
}
\label{subsec:final-evaluation}

After choosing $K$ in the dedicated scan in Sec.~\ref{subsec:model-selection} and the estimator in Sec.~\ref{subsec:mode-seeking-comparison}, we train \monster{} in ten new runs with different random seeds and all hyperparameters fixed.
For each run, we retain the checkpoint with the lowest validation \psixeight{} evaluated with the Most Probable Component estimator.
We then select the run with the lowest value of the same metric.
Finally, we report the test-set performance of the selected model and the mean and standard deviation across all ten runs.

\begin{table}
  \centering
  \small
  \caption{%
    Test-set momentum-error percentiles and root-mean-square errors (RMSEs).
    For \monster{} and \nuflows{}, each entry reports the result for the selected model, followed in parentheses by the mean $\pm$ sample standard deviation across ten training seeds.
    The kinematic reconstruction (Kin. Soln.) is deterministic, so no seed variation is reported.
    Lower values indicate better performance.
  }\label{tab:seed-variation-final-compact}
  \begin{tabular}{cccc}
    \toprule
    Metric & \monster{}  & \nuflows{}  & Kin. Soln. \\
    \midrule
    $P_{68}[\mathbf{p}]$ [GeV] & $\mathbf{59.7\;(60.8 \pm 0.6)}$ & $63.7\;(63.8 \pm 0.3)$ & $81.2$ \\
    $P_{95}[\mathbf{p}]$ [GeV] & $\mathbf{235.1\;(236.2 \pm 1.6)}$ & $250.2\;(251.3 \pm 1.1)$ & $297.1$ \\
    $\mathrm{RMSE}[E]$ [GeV] & $\mathbf{97.4\;(97.5 \pm 0.5)}$ & $101.0\;(101.1 \pm 0.4)$ & $108.1$ \\
    $\mathrm{RMSE}[p_T]$ [GeV] & $\mathbf{17.7\;(17.9 \pm 0.1)}$ & $18.5\;(18.6 \pm 0.1)$ & $22.7$ \\
    $\mathrm{RMSE}[\eta]$ & $\mathbf{1.053\;(1.056 \pm 0.007)}$ & $1.154\;(1.156 \pm 0.005)$ & $1.398$ \\
    $\mathrm{RMSE}[\phi]$ [rad] & $\mathbf{0.650\;(0.655 \pm 0.006)}$ & $0.669\;(0.671 \pm 0.002)$ & $0.737$ \\
    \bottomrule
  \end{tabular}%
\end{table}

Table~\ref{tab:seed-variation-final-compact} summarizes the test-set reconstruction performance and its variation across ten independent training seeds.
\monster{} achieves the lowest error for every metric considered, both for the selected model and for the seed-averaged results.
Relative to \nuflows{}, the mean errors are lower by approximately 5\% for \psixeight{}, 6\% for \pninefive{}, and 2--9\% for the individual kinematic root-mean-square errors (RMSEs).
The consistent ordering of the seed-averaged metrics shows that the improvement extends across the reconstructed neutrino kinematics.

The seed-to-seed standard deviations are generally somewhat larger for \monster{} than for \nuflows{}, indicating a modestly greater sensitivity to random initialization and stochastic training.
Nevertheless, these variations are small compared with the differences between the corresponding mean errors.
Thus, the improved reconstruction accuracy of \monster{} is maintained across training seeds, albeit with slightly greater variability than \nuflows{}.

\begin{figure}
  \begin{center}
    \includegraphics[width=0.45\textwidth]{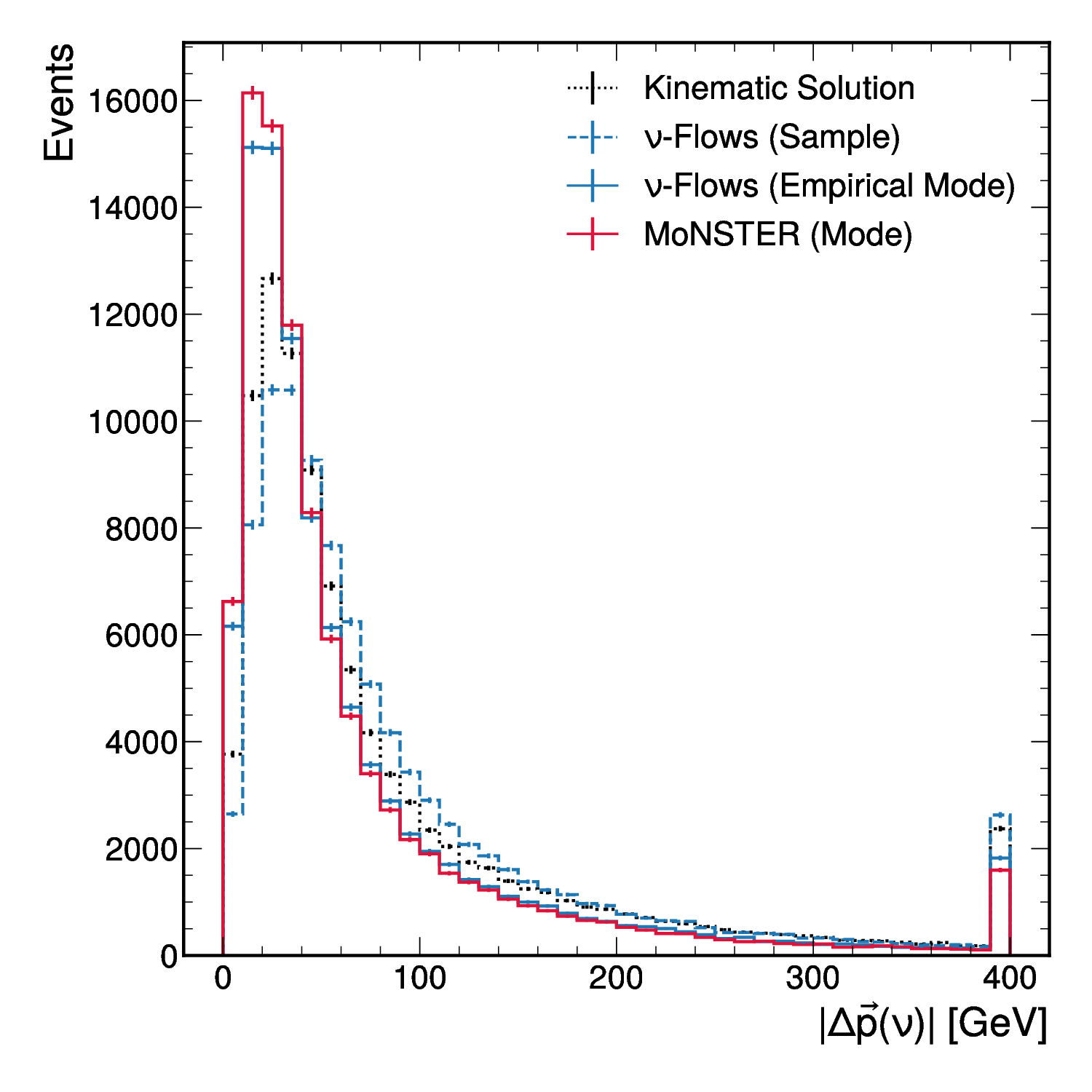}
  \end{center}
  \caption{%
    Distribution of the three-momentum residual $\deltap$ for the kinematic solution (black, dotted), \nuflowssample{} (blue, dashed), \nuflowsmode{} (blue, solid), and \monstermode{} (red, solid) on the test set.
    The last bin includes overflows.
  }
  \label{fig:dist-dp}
\end{figure}

Figure~\ref{fig:dist-dp} shows the distribution of $\deltap$ for the selected model on the test set, compared against the \nuflows{} baseline.
\monstermode{} yields smaller $\deltap$ than both \nuflowssample{} and \nuflowsmode{}, indicating better reconstruction accuracy.
The following subsections examine the selected model in more detail.

\subsection{Inference}
\label{subsec:inference}

\begin{figure}
  \begin{center}
    \includegraphics[width=0.32\textwidth]{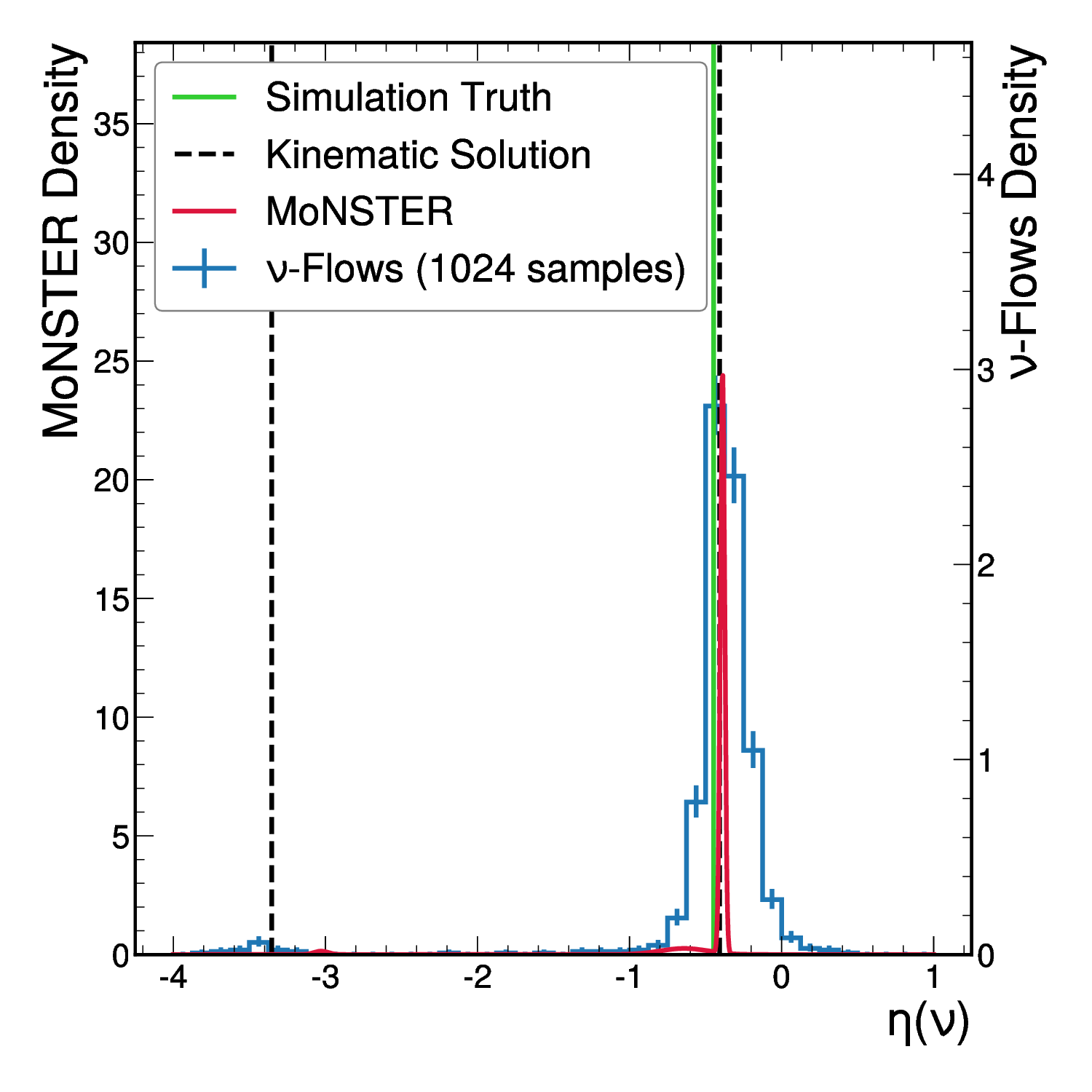}
    \includegraphics[width=0.32\textwidth]{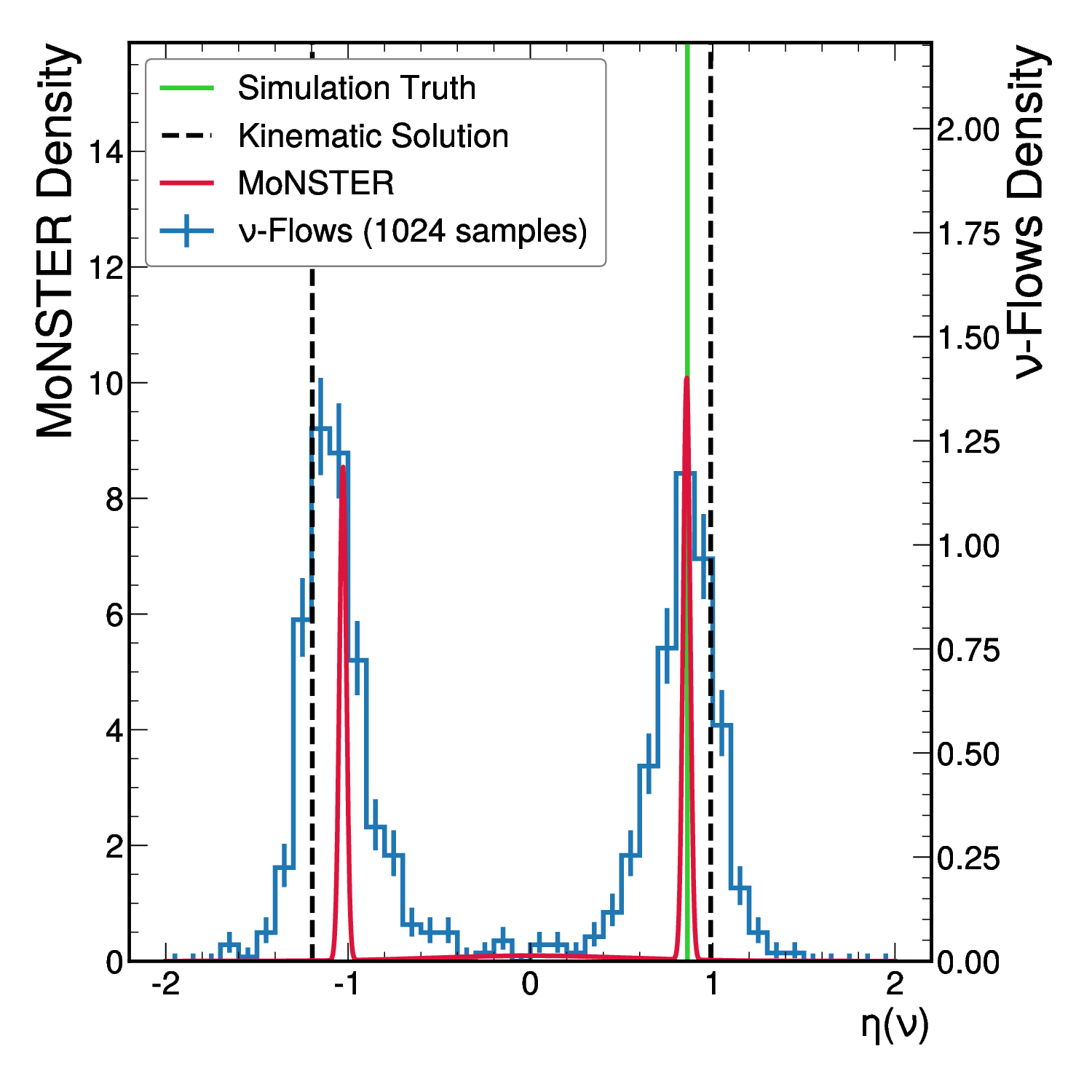}
    \includegraphics[width=0.32\textwidth]{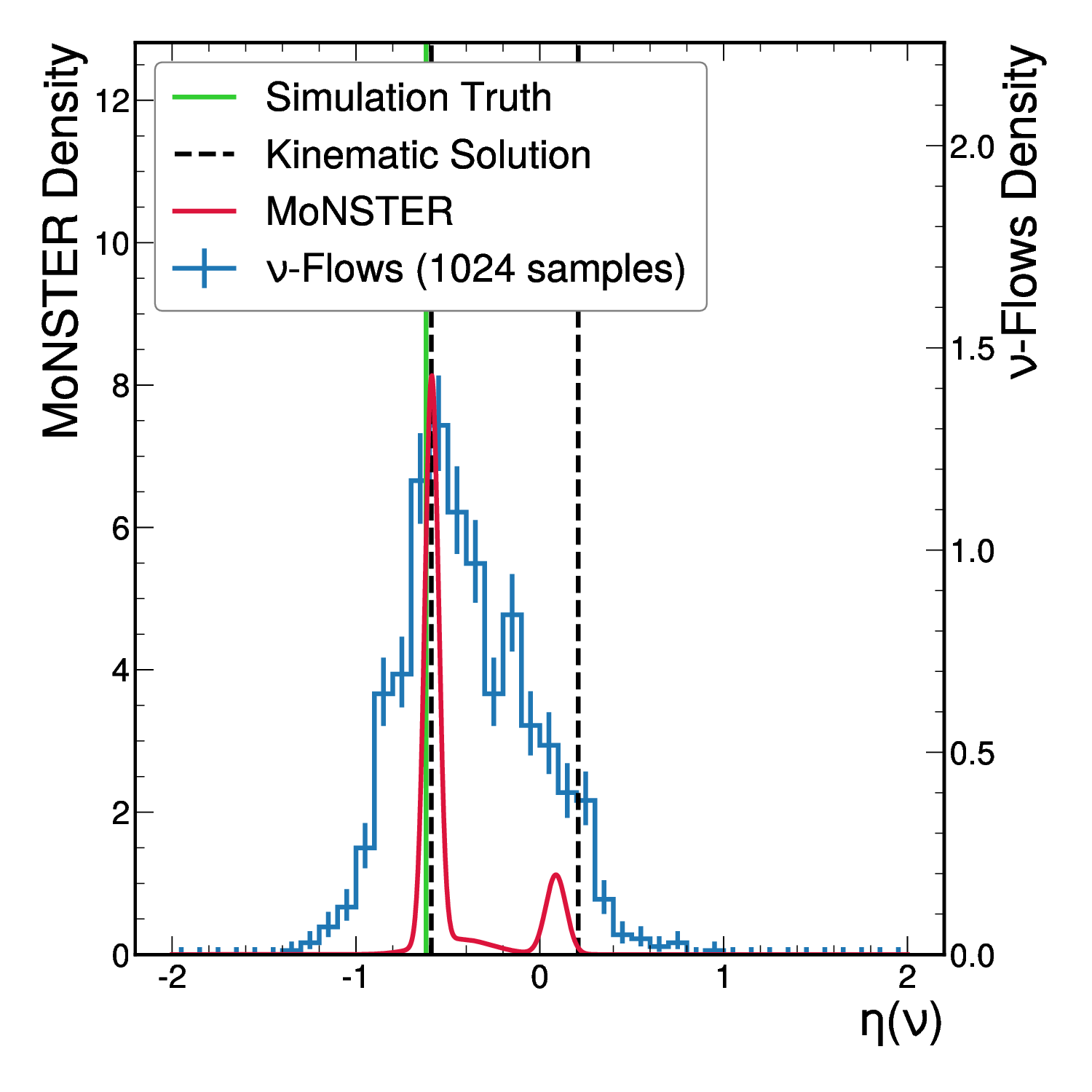}
  \end{center}
  \caption{
    Predicted marginal densities of the neutrino pseudorapidity $\eta(\nu)$ for three representative events from the test set.
    For \monster{}, the red curve is the exact $\eta(\nu)$ marginal of the predicted MVNMM.
    For \nuflows{}, the blue curve is estimated from 1024 samples drawn from the conditional flow.
    Vertical dashed black lines indicate the kinematic solutions implied by the $W$ boson mass constraint, and the vertical green line marks the truth value.
  }\label{fig:inference}
\end{figure}

Figure~\ref{fig:inference} compares the event-by-event conditional densities in $\eta(\nu)$ for three representative events from the test set, as predicted by \monster{} and \nuflows{}.
These examples span the range of structures encountered in semileptonic $t\bar{t}$ reconstruction, from clearly bimodal cases to nearly unimodal cases.
For \monster{}, the displayed curve is obtained by analytically marginalizing the predicted MVNMM over the other two momentum components, which yields a one-dimensional normal mixture.
For \nuflows{}, the corresponding marginal is estimated from 1024 samples drawn from the conditional flow.
The truth value is shown as a vertical green line.
Across these examples, \monster{} and \nuflows{} identify similar high-density regions and reproduce both the well-separated and nearly merged two-branch configurations.
In each case, the truth value lies in or near a region of substantial predicted density.

\subsection{Performance versus Sample Size}
\label{subsec:metric-vs-sampling-size}

\begin{figure}
  \begin{center}
    \includegraphics[width=0.32\textwidth]{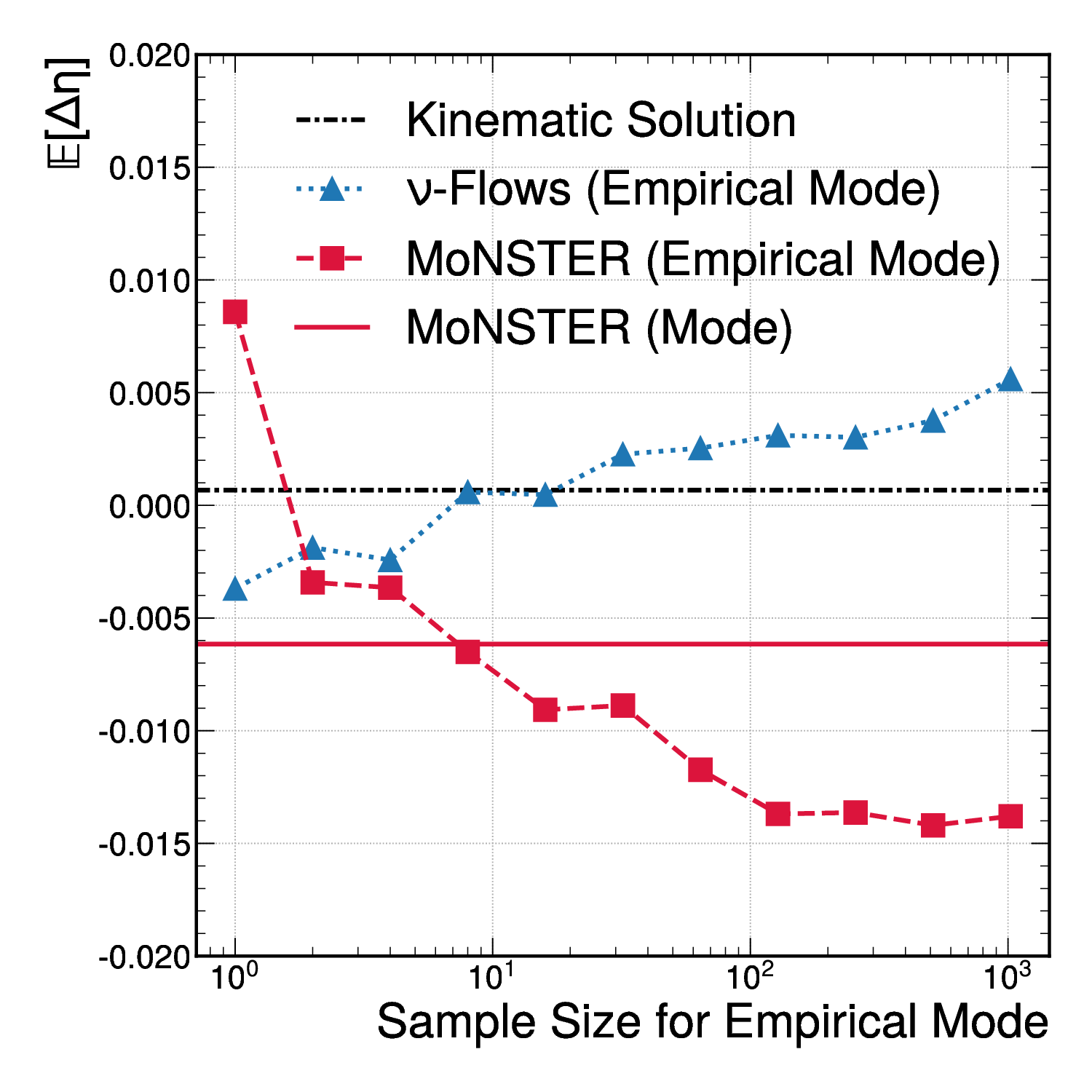}
    \includegraphics[width=0.32\textwidth]{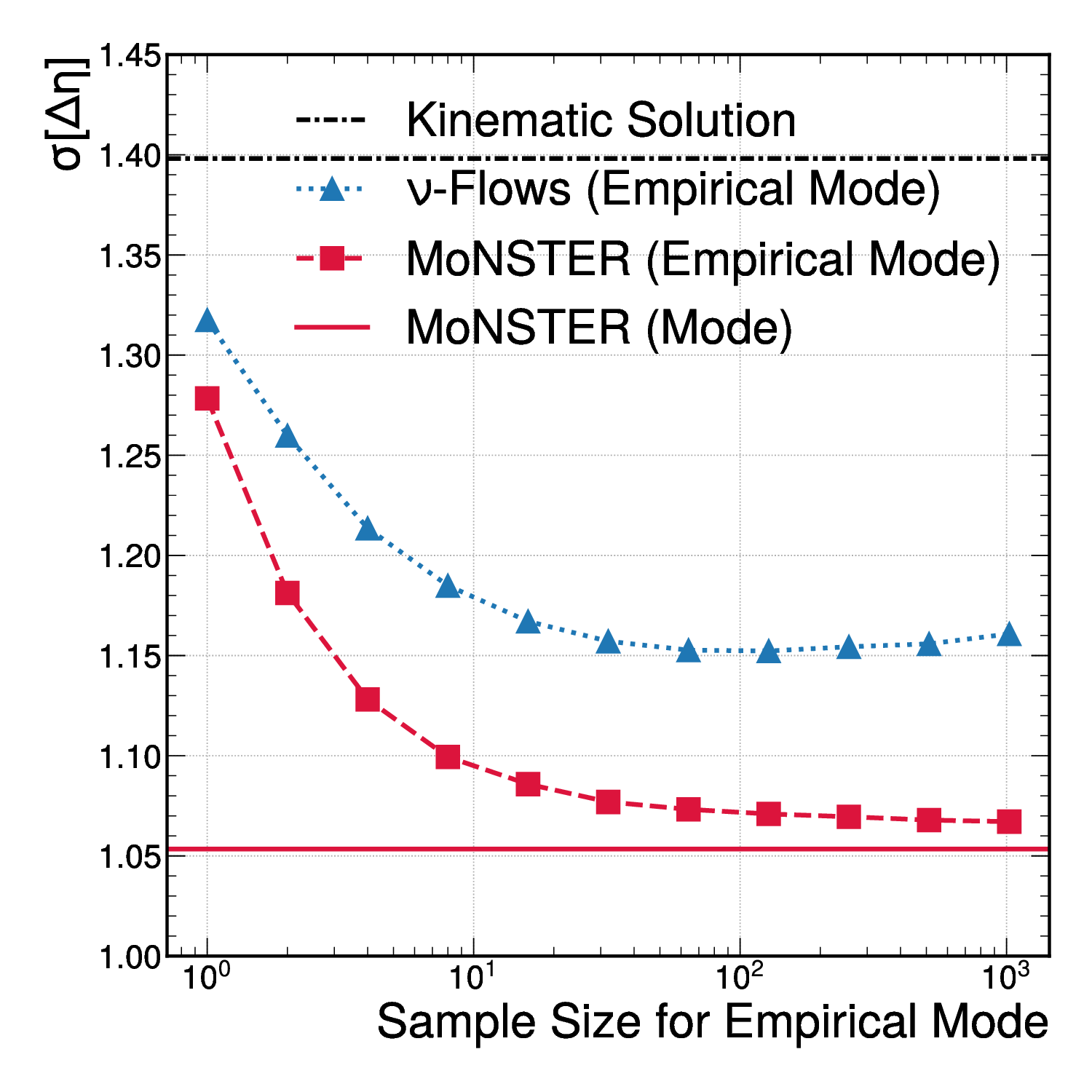}
    \includegraphics[width=0.32\textwidth]{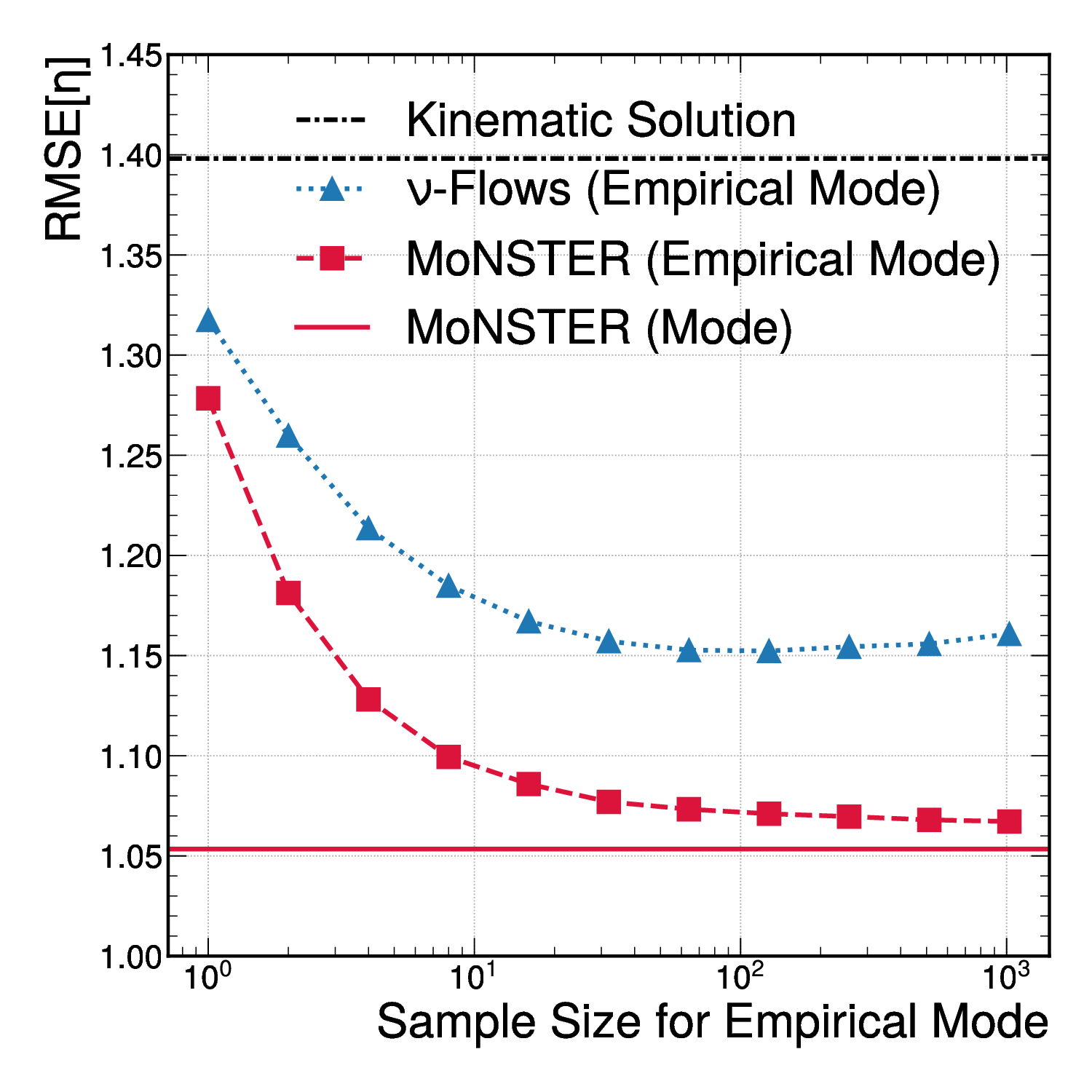}
  \end{center}
  \caption{
    Bias (left), resolution (center), and RMSE (right) of the reconstructed neutrino pseudorapidity $\eta(\nu)$ as a function of the number of samples used to estimate the empirical mode, for \nuflowsmode{} (blue triangles) and \monsterempmode{} (red squares).
    Results are evaluated on the validation set using the selected model for each method.
    The horizontal red solid line indicates the performance of \monstermode{}, which is obtained directly from the predicted mixture parameters and requires no sampling.
    The black dash-dotted line indicates the kinematic solution.
  }\label{fig:metric-vs-sampling-size}
\end{figure}

Figure~\ref{fig:metric-vs-sampling-size} shows the bias, resolution, and RMSE of the reconstructed neutrino pseudorapidity $\eta(\nu)$ as a function of the number of samples used to estimate the empirical mode, for both \nuflowsmode{} and \monsterempmode{}.
The empirical-mode estimators draw a fixed number of samples per event from the corresponding conditional density and retain the sample with the highest predicted density.

For both methods, the resolution and RMSE decrease as the sample size increases, reflecting a more accurate approximation of the mode as more candidates are evaluated.
At a given sample size, \monsterempmode{} tends to improve resolution and RMSE relative to \nuflowsmode{}, suggesting that the mixture density network provides a somewhat more accurate mode estimate for the same number of samples.
The horizontal solid line in each panel indicates the performance of \monstermode{}, obtained directly from the closed-form MVNMM density, which requires no stochastic sampling.

\subsection{Reconstructed Distributions}
\label{subsec:dist}

\begin{figure}
  \begin{center}
    \includegraphics[width=0.45\textwidth]{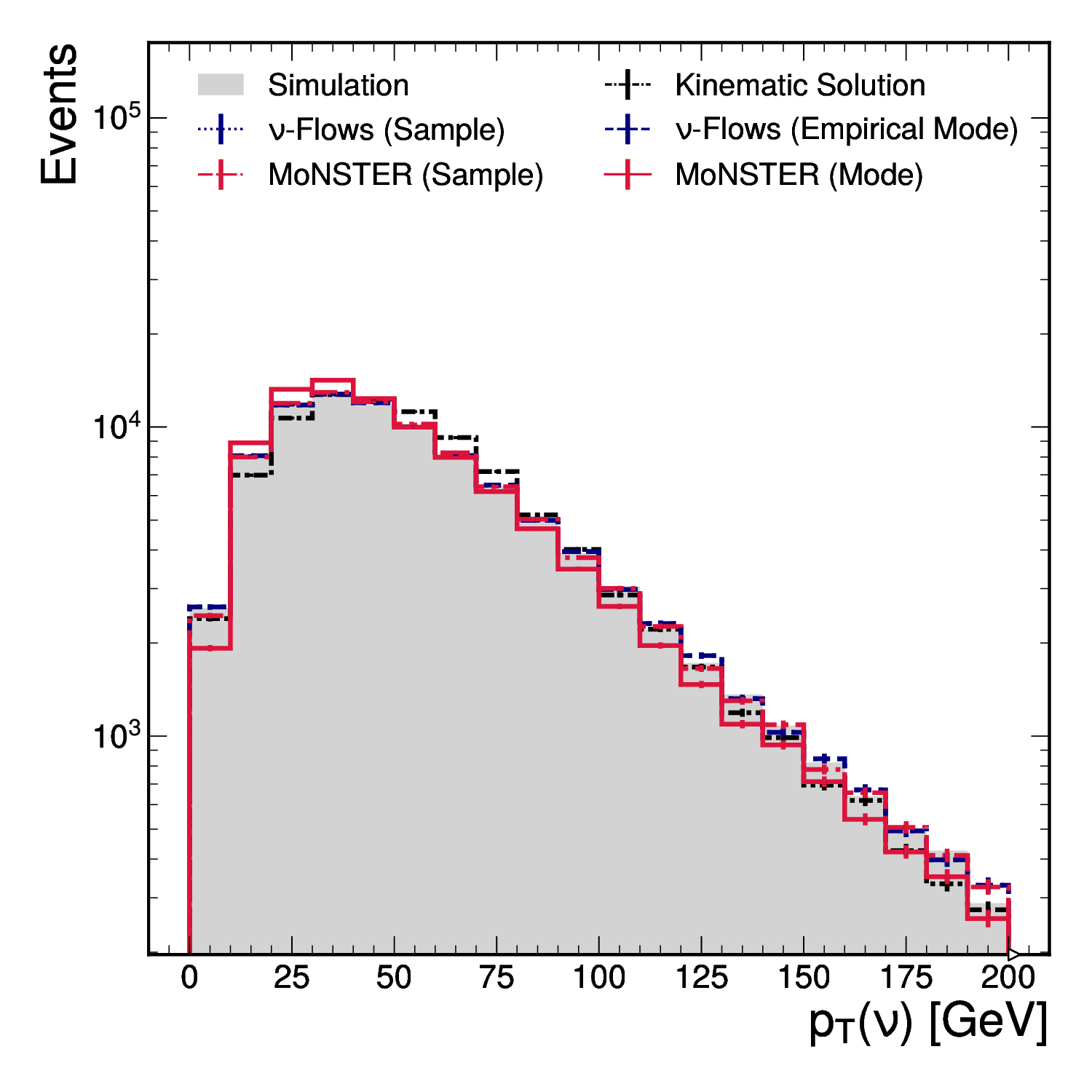}
    \includegraphics[width=0.45\textwidth]{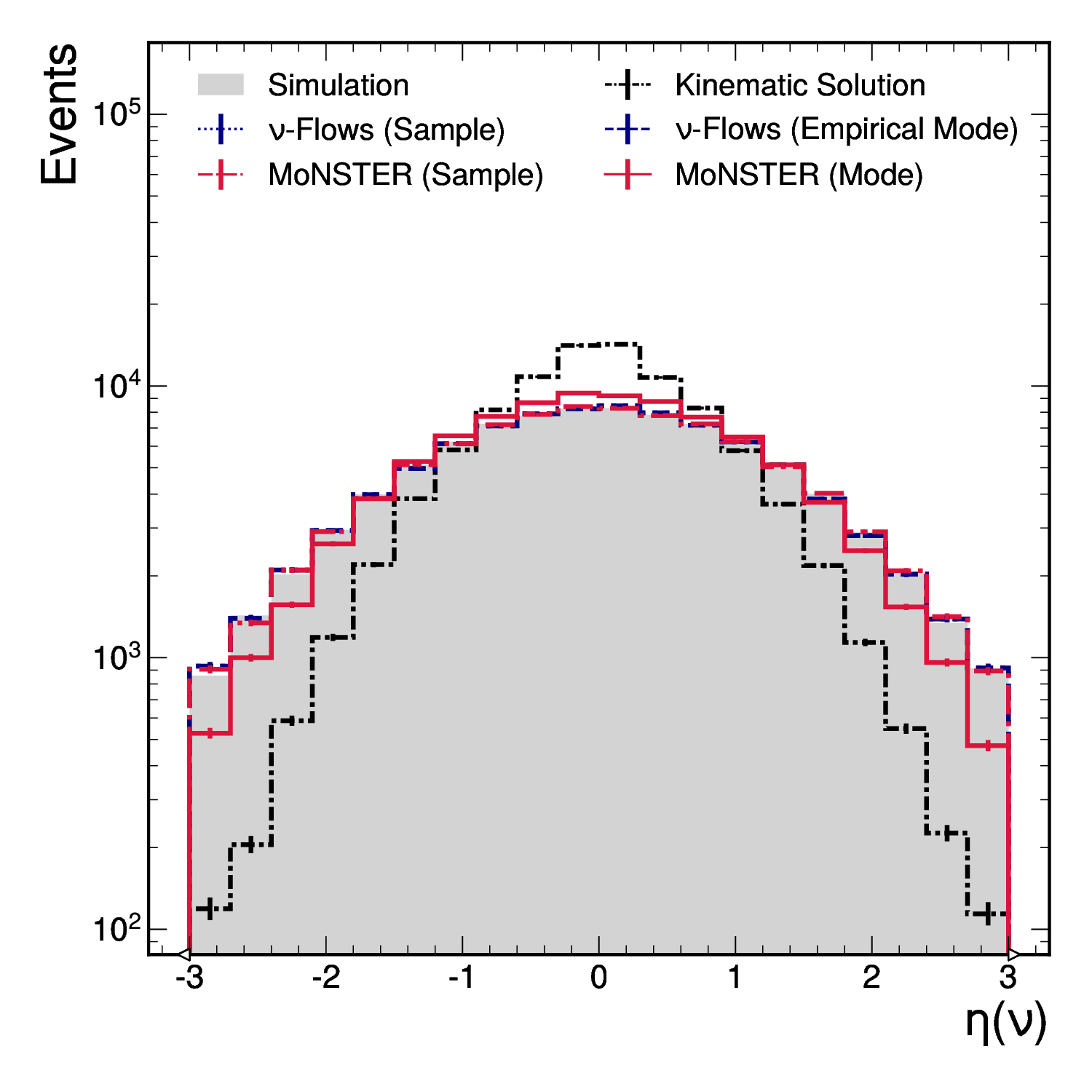}
    \includegraphics[width=0.45\textwidth]{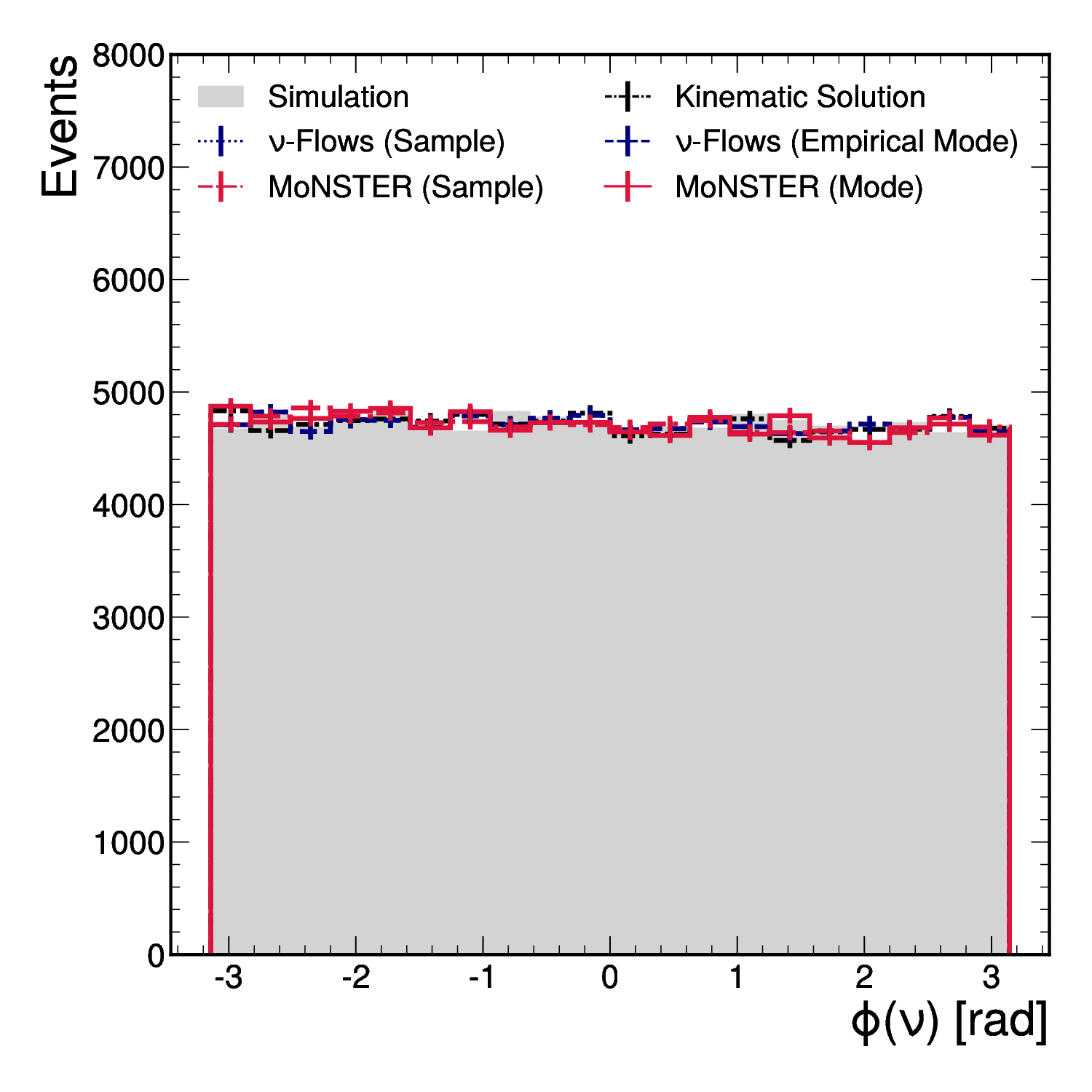}
    \includegraphics[width=0.45\textwidth]{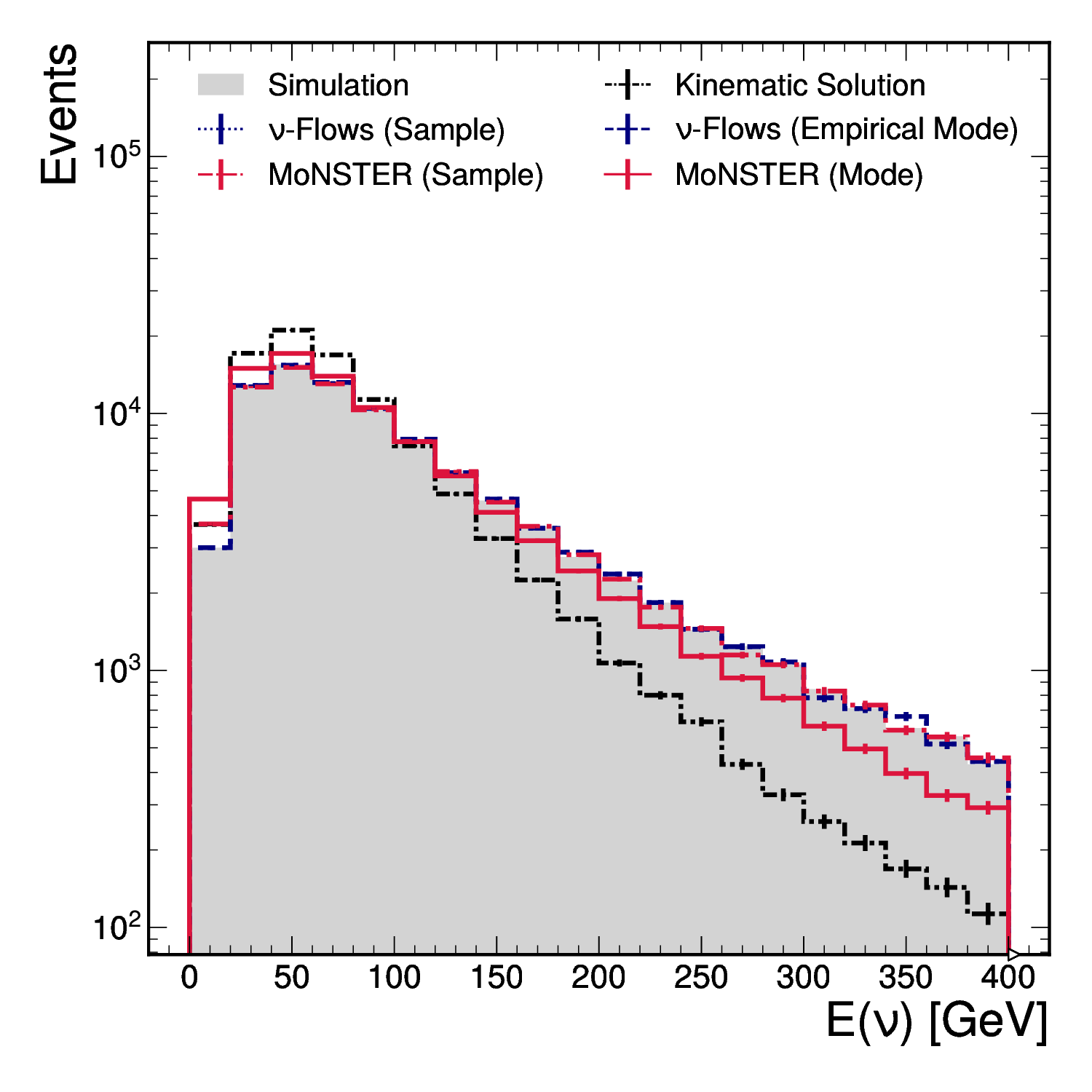}
  \end{center}
  \caption{
    Distributions of reconstructed neutrino kinematic variables: transverse momentum $p_{T}(\nu)$ (upper left), pseudorapidity $\eta(\nu)$ (upper right), azimuthal angle $\phi(\nu)$ (lower left), and energy $E(\nu)$ (lower right).
    The gray-filled histogram shows the truth-level distribution from simulation.
    Colored markers show the point estimates from \nuflowssample{}, \nuflowsmode{}, \monstersample{}, \monstermode{}, and the kinematic solution.
  }\label{fig:distributions}
\end{figure}

Figure~\ref{fig:distributions} compares the distributions of the reconstructed neutrino kinematic variables $p_{T}(\nu)$, $\eta(\nu)$, $\phi(\nu)$, and $E(\nu)$ with the corresponding truth-level distributions from simulation.
For the sample-based estimators, \nuflowssample{} and \monstersample{}, agreement with the truth histograms is expected if the learned conditional density approximates the true conditional distribution $p(\mathbf{y} \mid \mathbf{x})$.
Drawing one sample per event and aggregating over events then emulates the shape of the corresponding truth-level marginal distribution in expectation.
This is therefore a population-level consistency check, not a stringent test of event-by-event reconstruction accuracy, and it should not be used to rank the methods.

For the mode-based estimators, the situation is different.
They return deterministic high-density representatives of the learned conditional density, so their aggregated one-dimensional distributions need not coincide with the truth-level marginals even when the underlying conditional density is well modeled.
We therefore use Fig.~\ref{fig:distributions} only as a qualitative check that the inferred point estimates do not introduce obvious distortions in the overall kinematic spectra.

\subsection{Performance as a Function of Kinematic Variables}
\label{subsec:binned-metric}

\begin{figure}
  \begin{center}
    \includegraphics[width=0.32\textwidth]{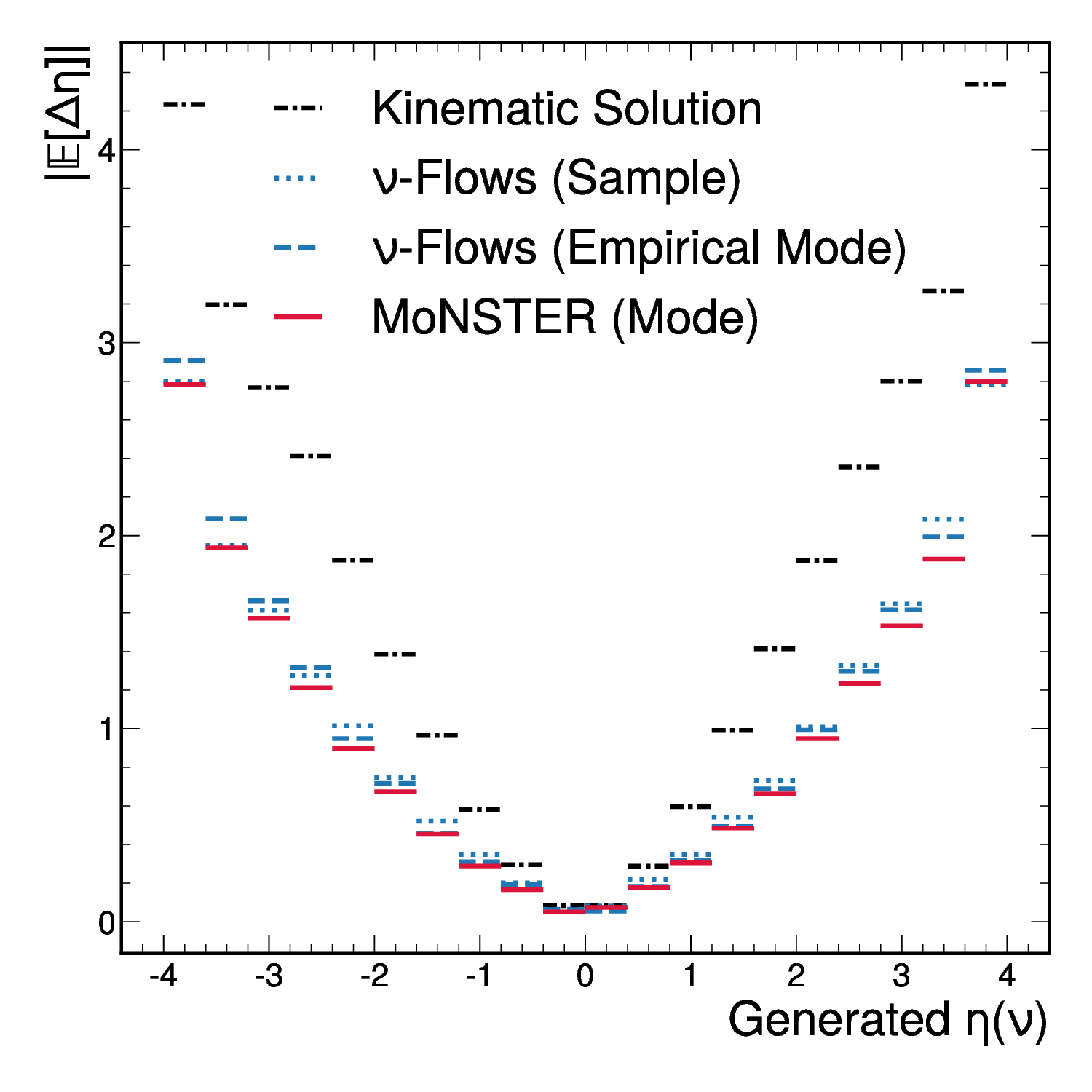}
    \includegraphics[width=0.32\textwidth]{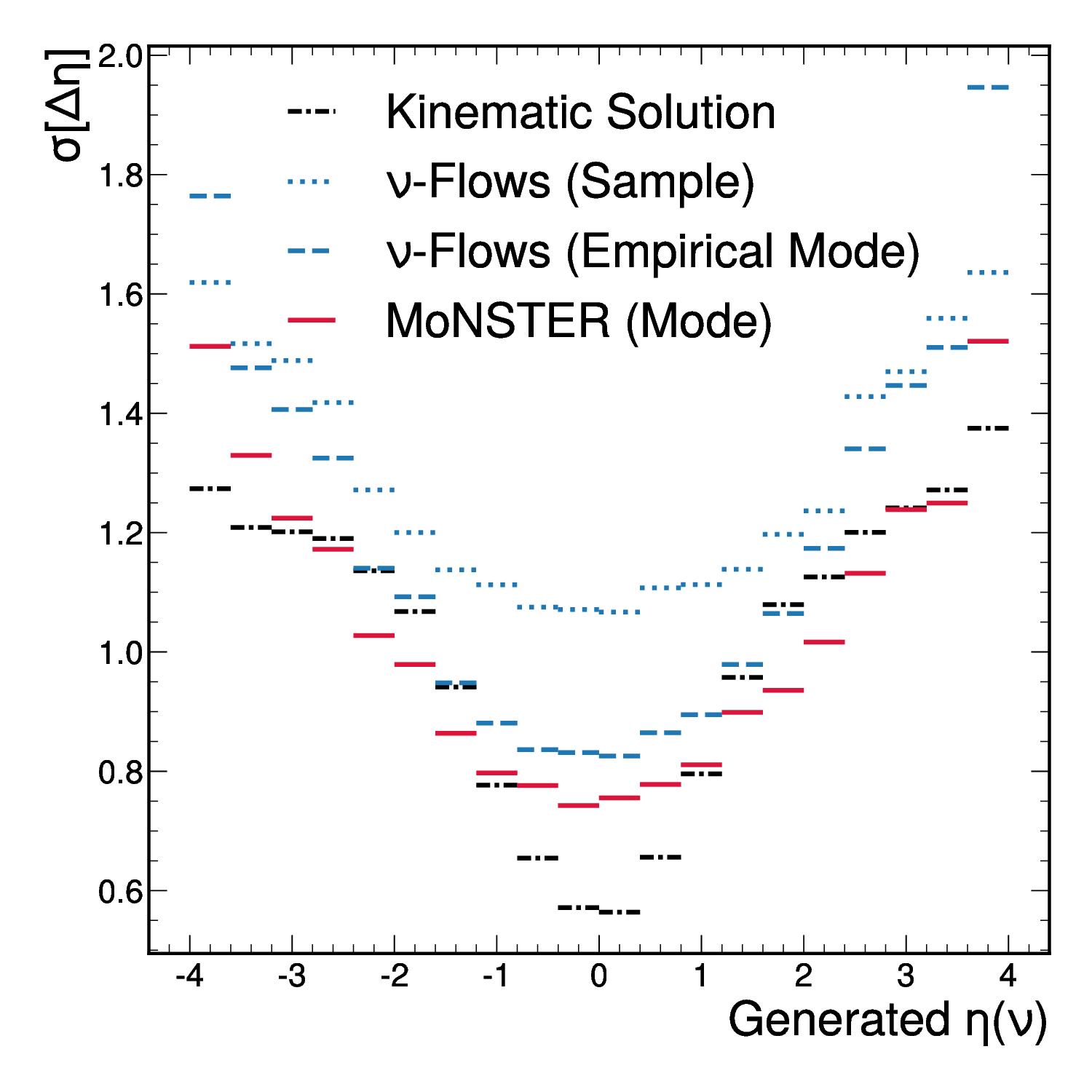}
    \includegraphics[width=0.32\textwidth]{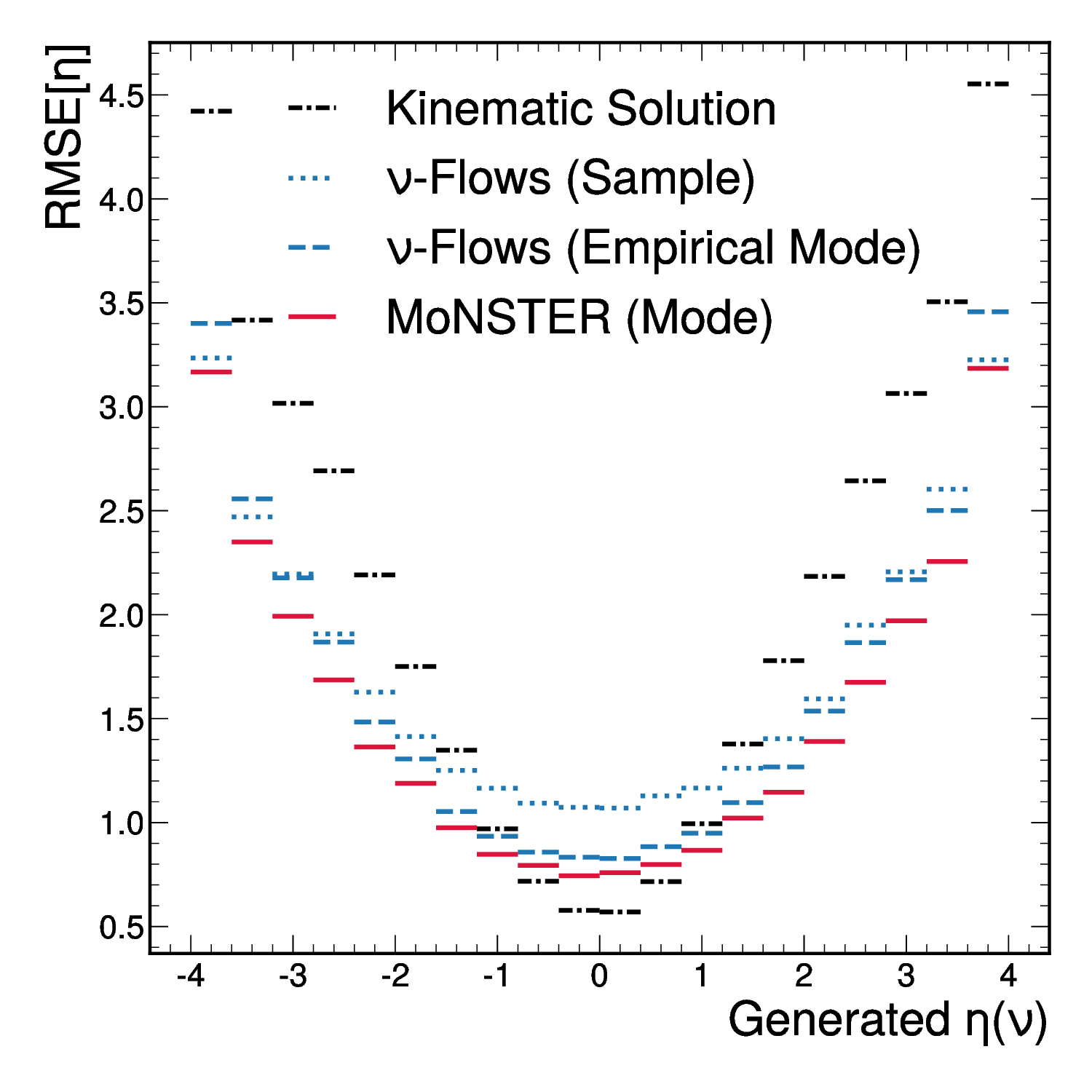}
  \end{center}
  \caption{
    Absolute bias (left), resolution (center), and RMSE (right) of the reconstructed neutrino pseudorapidity $\eta(\nu)$, shown as a function of the generated $\eta(\nu)$.
    Results are shown for the kinematic solution (black dash-dotted), \nuflowssample{} (blue dotted), \nuflowsmode{} (blue dashed), and \monstermode{} (red solid).
  }\label{fig:binned-metric-eta}
\end{figure}

\begin{figure}
  \begin{center}
    \includegraphics[width=0.32\textwidth]{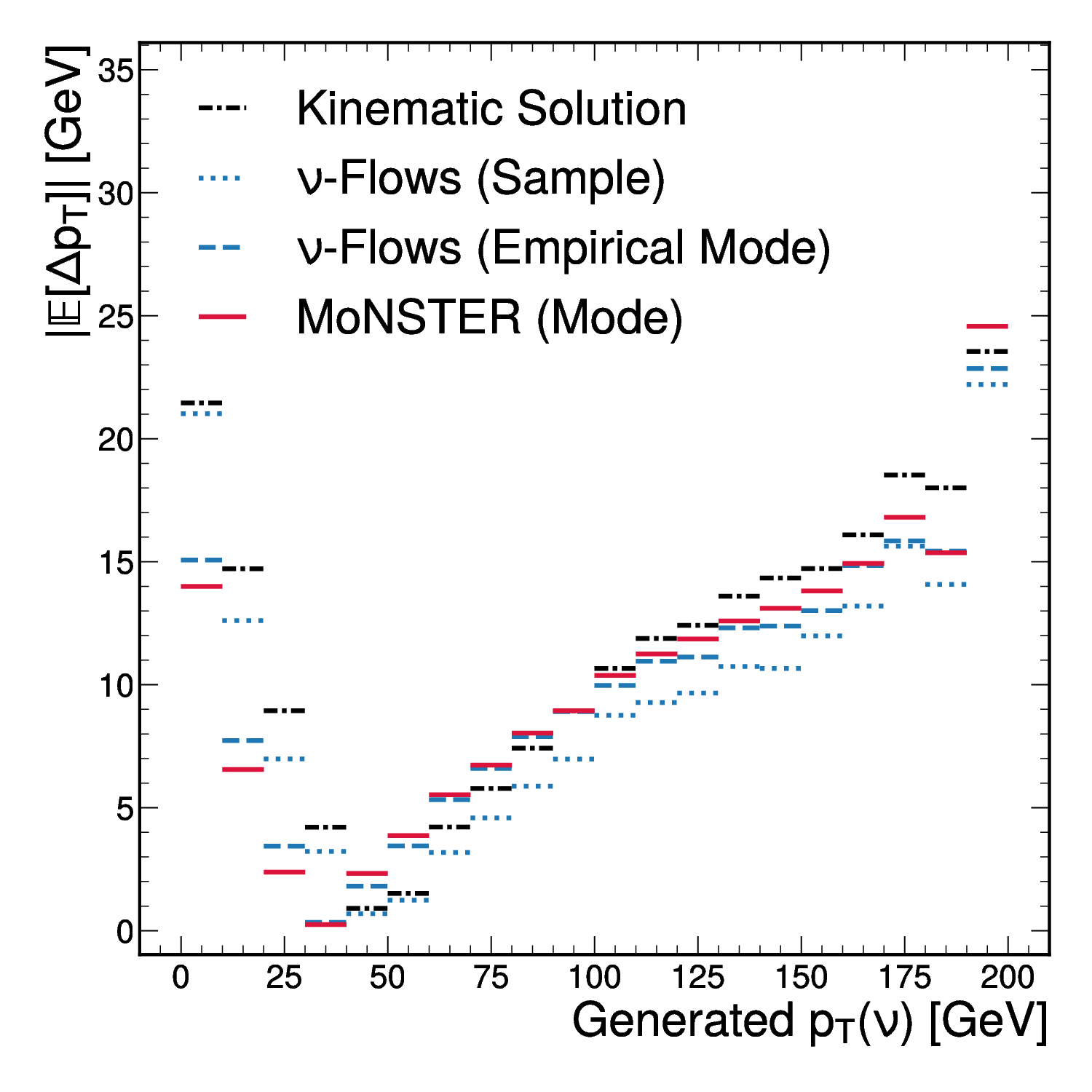}
    \includegraphics[width=0.32\textwidth]{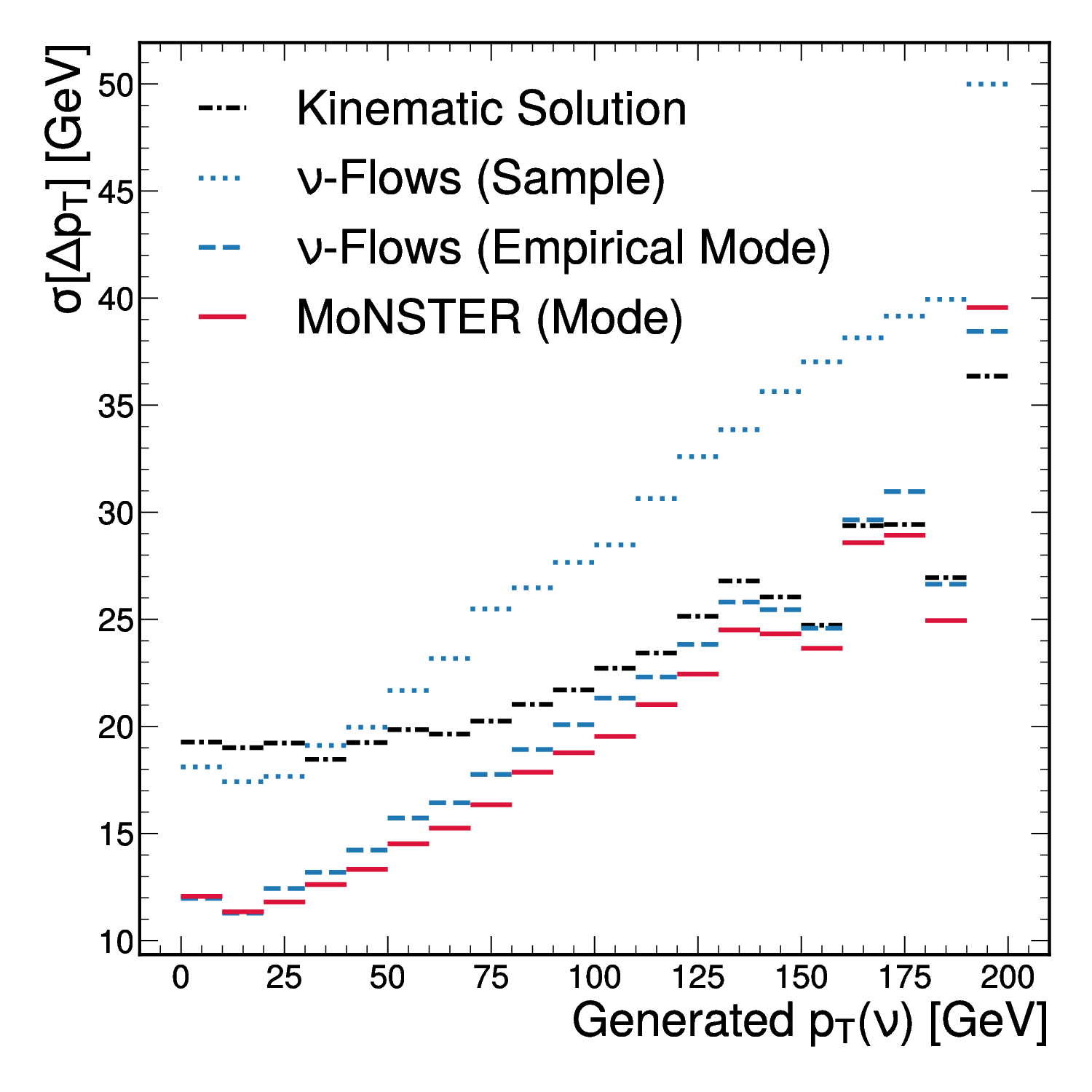}
    \includegraphics[width=0.32\textwidth]{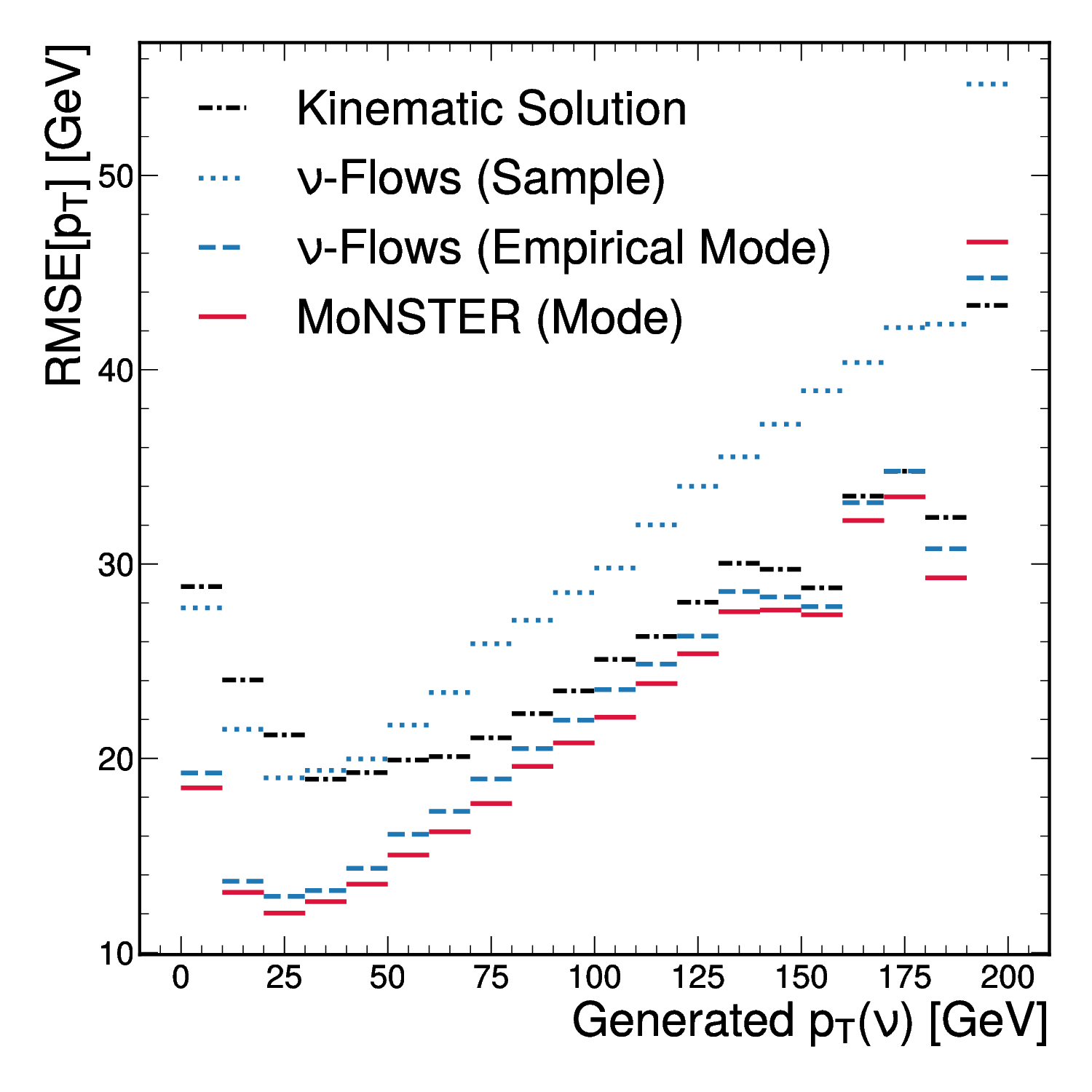}
  \end{center}
  \caption{
    Absolute bias (left), resolution (center), and RMSE (right) of the reconstructed neutrino transverse momentum $p_T(\nu)$, shown as a function of the generated $p_T(\nu)$ in bins of equal width.
    Results are shown for the kinematic solution (black dash-dotted), \nuflowssample{} (blue dotted), \nuflowsmode{} (blue dashed), and \monstermode{} (red solid).
  }\label{fig:binned-metric-pt}
\end{figure}

\begin{figure}
  \begin{center}
    \includegraphics[width=0.32\textwidth]{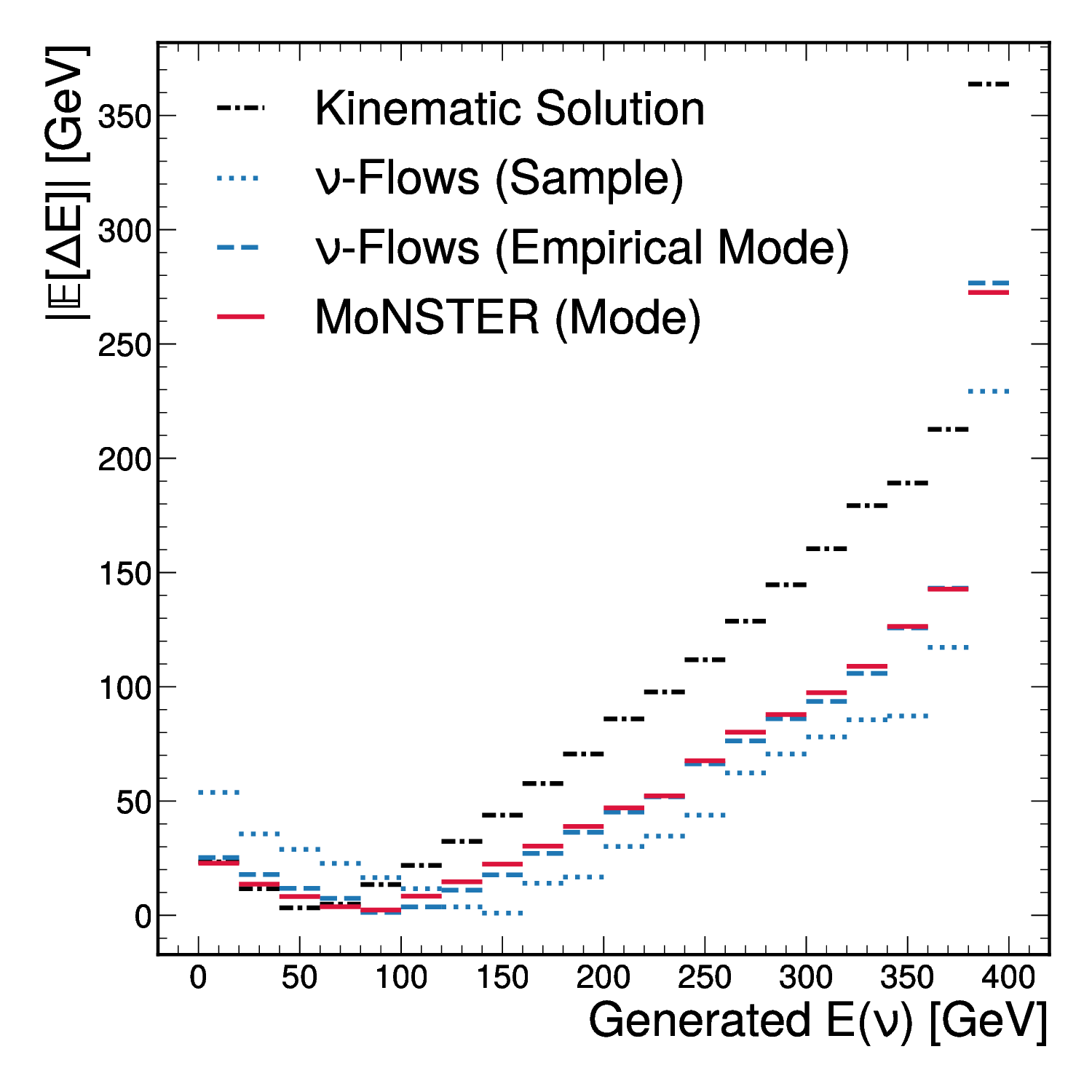}
    \includegraphics[width=0.32\textwidth]{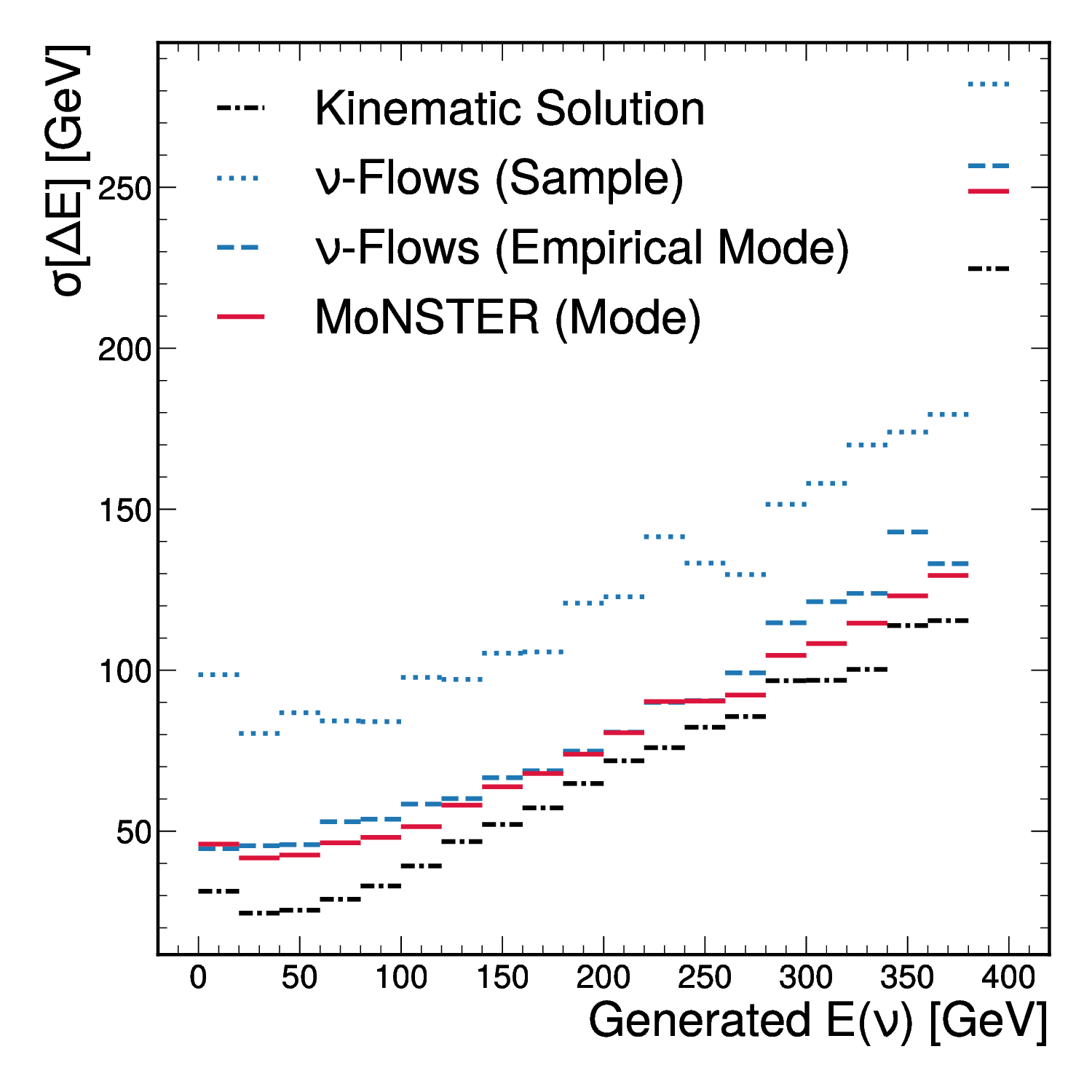}
    \includegraphics[width=0.32\textwidth]{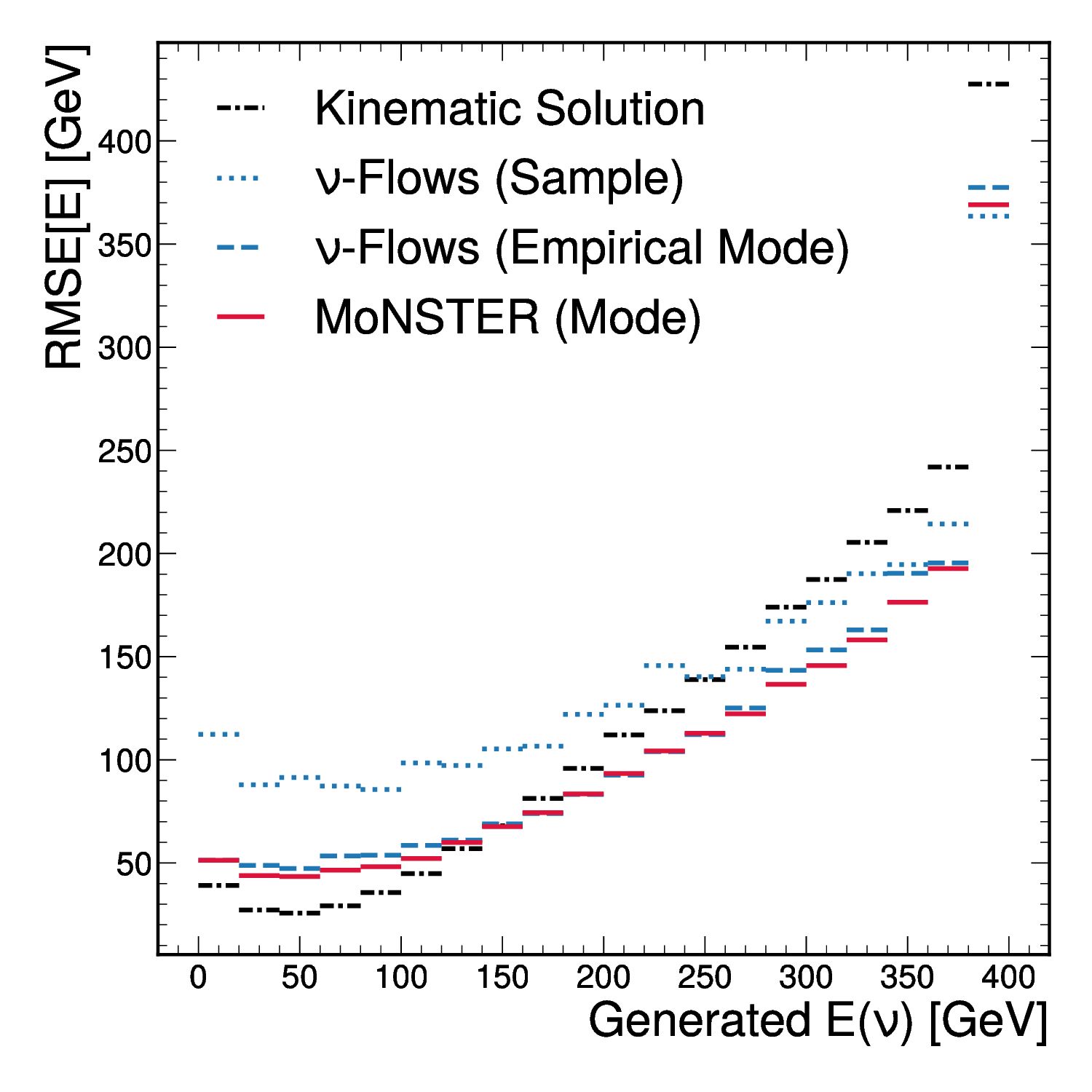}
  \end{center}
  \caption{
    Absolute bias (left), resolution (center), and RMSE (right) of the reconstructed neutrino energy $E(\nu)$, shown as a function of the generated $E(\nu)$ in GeV.
    Results are shown for the kinematic solution (black dash-dotted), \nuflowssample{} (blue dotted), \nuflowsmode{} (blue dashed), and \monstermode{} (red solid).
  }\label{fig:binned-metric-energy}
\end{figure}

Figures~\ref{fig:binned-metric-eta}, \ref{fig:binned-metric-pt}, and~\ref{fig:binned-metric-energy} show the absolute bias, resolution, and RMSE of the reconstructed neutrino pseudorapidity $\eta(\nu)$, transverse momentum $p_T(\nu)$, and energy $E(\nu)$ as functions of the corresponding generator-level quantities.
Across these observables, \monstermode{} generally improves resolution and RMSE relative to \nuflowsmode{}, while maintaining a comparable absolute bias.
The size of the improvement is not uniform across phase space, so the comparison is most meaningful on a bin-by-bin basis.

For $\eta(\nu)$, shown in Fig.~\ref{fig:binned-metric-eta}, all three metrics are approximately symmetric about $\eta(\nu)=0$.
The metrics exhibit a characteristic valley at central pseudorapidity and increase toward the forward region.
The reconstruction is most accurate in the central region, which is also the high-statistics region, and degrades toward larger $|\eta(\nu)|$.
Over most of the $\eta$ range, \monstermode{} remains below \nuflowsmode{} in resolution and RMSE, with a similar or slightly smaller absolute bias.

Figures~\ref{fig:binned-metric-pt} and~\ref{fig:binned-metric-energy} exhibit closely related behavior.
For both $p_T(\nu)$ and $E(\nu)$, the absolute bias and RMSE initially decrease toward a minimum and then increase, with the largest errors within the plotted range occurring at the high end.
The decrease in RMSE at low energy is shallower than at low $p_T$.
The qualitative similarity between the $p_{T}(\nu)$ and $E(\nu)$ dependences is expected, since these observables are strongly correlated for a nearly massless neutrino.

The resolution follows the same broad pattern, although the details differ slightly between the two observables.
In Fig.~\ref{fig:binned-metric-pt}, the $p_T(\nu)$ resolution shows a shallow minimum at low-to-intermediate $p_T$ and then worsens steadily toward high $p_T$.
In Fig.~\ref{fig:binned-metric-energy}, the $E(\nu)$ resolution is comparatively flat over the low-energy region before rising at higher energy.
Overall, the figures indicate that the main advantage of \monstermode{} is a systematic reduction in spread, rather than a large change in central-value bias.


\section{Conclusion}
\label{sec:conclusion}

We introduce \monster{}, a mixture density network for neutrino reconstruction at hadron colliders.
\monster{} models the conditional neutrino momentum distribution with a multivariate normal mixture and uses the closed-form density for sampling-free mode estimation.
On the public semileptonic $t\bar{t}$ benchmark, \monstermode{} achieves better resolution and RMSE than \nuflowsmode{} across the kinematic variables considered, while maintaining comparable bias.
Notably, \monster{} attains this improvement while using strictly fewer inputs than the baseline, omitting the jet and $b$-tagged jet multiplicities that \nuflows{} relies on.
The closed-form mixture density additionally permits sampling-free mode estimation at inference time, in contrast to the stochastic candidate search required by flow-based methods.
Overall, these results indicate that, for a low-dimensional target with moderate multimodality such as the neutrino momentum, an explicit mixture density is sufficient to outperform a normalizing flow on this benchmark at a lower inference cost.

As the present study is limited to the single-neutrino final state of semileptonic $t\bar{t}$ events, extending the mixture density approach to multi-neutrino topologies is a natural direction for future work.
Reconstruction in the forward and high-$p_{T}$ regions, where all methods degrade, also remains an open target for improvement.

\begin{acknowledgments}
S.Y. and J.L. are supported by the National Research Foundation of Korea (NRF) grant No. 2023R1A2C2002751.
S.Y., J.L., and J.G. are supported by the NRF grant No. RS-2008-NR007227.
This work was supported by the Yonsei Fellowship, funded by Lee Youn Jae.
This research was supported by the Yonsei University Research Fund (2026-22-0257).

\end{acknowledgments}

\bibliography{ref}

@techreport{bishop1994mixture,
author = {Bishop, Christopher M},
title = {Mixture Density Networks},
year = {1994},
month = {January},
institution = {Aston University},
url = {https://www.microsoft.com/en-us/research/publication/mixture-density-networks/},
number = {NCRG/94/004},
}

@article{Graziani:2021vai,
    author = "Graziani, Giacomo and Anderlini, Lucio and Mariani, Saverio and Franzoso, Edoardo and Pappalardo, Luciano Libero and di Nezza, Pasquale",
    title = "{A Neural-Network-defined Gaussian Mixture Model for particle identification applied to the LHCb fixed-target programme}",
    eprint = "2110.10259",
    archivePrefix = "arXiv",
    primaryClass = "hep-ex",
    reportNumber = "LHCb-DP-2021-007",
    doi = "10.1088/1748-0221/17/02/P02018",
    journal = "{JINST}",
    volume = "17",
    number = "02",
    pages = "P02018",
    year = "2022"
}

@article{ATLAS:2023zca,
    author = "Aad, Georges and others",
    collaboration = "ATLAS",
    title = "{Simultaneous energy and mass calibration of large-radius jets with the ATLAS detector using a deep neural network}",
    eprint = "2311.08885",
    archivePrefix = "arXiv",
    primaryClass = "hep-ex",
    reportNumber = "CERN-EP-2023-250",
    doi = "10.1088/2632-2153/ad611e",
    journal = "Mach. Learn. Sci. Tech.",
    volume = "5",
    number = "3",
    pages = "035051",
    year = "2024"
}

@misc{Das:2026mff,
    author = "Das, Arghya Ranjan and others",
    title = "{On-chip probabilistic inference for charged-particle tracking at the sensor edge}",
    eprint = "2602.15946",
    archivePrefix = "arXiv",
    primaryClass = "physics.ins-det",
    reportNumber = "FERMILAB-PUB-26-0100-CSAID-ETD",
    month = "2",
    year = "2026"
}

@article{Khoda:2687968,
    author = "Khoda, Elham E.",
    editor = "Roig Garc{\'e}s, Pablo and Bautista Guzman, Irais and Fern{\'a}ndez T{\'e}llez, Arturo and Mart{\'\i}nez Hern{\'a}ndez, Mario Iv{\'a}n",
    collaboration = "ATLAS",
    title = "{ATLAS pixel cluster splitting using Mixture Density Networks}",
    reportNumber = "ATL-PHYS-PROC-2019-082",
    doi = "10.22323/1.350.0009",
    journal = "PoS",
    volume = "LHCP2019",
    pages = "009",
    year = "2019"
}

@techreport{ATL-PHYS-PUB-2018-014,
      author = "ATLAS",
      title         = "{Performance of mass-decorrelated jet substructure
                       observables for hadronic two-body decay tagging in ATLAS}",
      institution   = "CERN",
      number  = "ATL-PHYS-PUB-2018-014",
      address       = "Geneva",
      year          = "2018",
      url           = "https://cds.cern.ch/record/2630973",
}

@techreport{ATL-PHYS-PUB-2022-033,
      author = "ATLAS",
      title = "{Performance of ATLAS Pixel Detector and Track Reconstruction at the start of Run~3 in LHC Collisions at $\sqrt{s}=900$ GeV}",
      institution   = "CERN",
      number  = "ATL-PHYS-PUB-2022-033",
      address       = "Geneva",
      year          = "2022",
      url           = "https://cds.cern.ch/record/2814766",
}

@article{Leigh:2022lpn,
    author = "Leigh, Matthew and Raine, John Andrew and Zoch, Knut and Golling, Tobias",
    title = "{$\nu$-Flows: Conditional Neutrino Regression}",
    eprint = "2207.00664",
    archivePrefix = "arXiv",
    primaryClass = "hep-ph",
    doi = "10.21468/SciPostPhys.14.6.159",
    journal = "SciPost Phys.",
    volume = "14",
    number = "6",
    pages = "159",
    year = "2023"
}

@misc{zoch_2022_6782987,
  author       = {Zoch, Knut and
                  Raine, John Andrew and
                  Ehrke, Lukas and
                  Sengupta, Debajyoti and
                  Leigh, Matthew and
                  Golling, Tobias},
  title        = {Semileptonic ttbar Neutrino Regression Dataset},
  month        = jun,
  year         = 2022,
  publisher    = {Zenodo},
  version      = 2,
  doi          = {10.5281/zenodo.6782987},
  url          = {https://doi.org/10.5281/zenodo.6782987},
  howpublished = {\url{https://doi.org/10.5281/zenodo.6782987}},  
}

@misc{Leigh_neutrino_flows,
  author       = {Matthew Leigh},
  title        = "{neutrino\_flows}",
  url = {https://github.com/mattcleigh/neutrino_flows},
  note         = {accessed 2026-03-30},
  year         = {2022},
  howpublished = {\url{https://github.com/mattcleigh/neutrino_flows}},  
}

@article{Raine:2023fko,
    author = "Raine, John Andrew and Leigh, Matthew and Zoch, Knut and Golling, Tobias",
    title = "{Fast and improved neutrino reconstruction in multineutrino final states with conditional normalizing flows}",
    eprint = "2307.02405",
    archivePrefix = "arXiv",
    primaryClass = "hep-ph",
    doi = "10.1103/PhysRevD.109.012005",
    journal = "Phys. Rev. D",
    volume = "109",
    number = "1",
    pages = "012005",
    year = "2024"
}

@misc{DBLP:journals/corr/abs-1907-02392,
      title="{Guided Image Generation with Conditional Invertible Neural Networks}", 
      author={Lynton Ardizzone and Carsten Lüth and Jakob Kruse and Carsten Rother and Ullrich Köthe},
      year={2019},
      eprint={1907.02392},
      archivePrefix={arXiv},
      primaryClass={cs.CV},
      url={https://arxiv.org/abs/1907.02392}, 
}

@misc{DBLP:journals/corr/ZaheerKRPSS17,
      title="{Deep Sets}", 
      author={Manzil Zaheer and Satwik Kottur and Siamak Ravanbakhsh and Barnabas Póczos and Ruslan Salakhutdinov and Alexander Smola},
      year={2018},
      eprint={1703.06114},
      archivePrefix={arXiv},
      primaryClass={cs.LG},
      url={https://arxiv.org/abs/1703.06114}, 
}

@inproceedings{DBLP:conf/nips/DurkanB0P19,
 author = {Durkan, Conor and Bekasov, Artur and Murray, Iain and Papamakarios, George},
 booktitle = "{Advances in Neural Information Processing Systems}",
 pages = {},
 publisher = {Curran Associates, Inc.},
 title = "{Neural Spline Flows}",
 url = {https://proceedings.neurips.cc/paper_files/paper/2019/file/7ac71d433f282034e088473244df8c02-Paper.pdf},
 volume = {32},
 year = {2019}
}

@inproceedings{NEURIPS2018_d139db6a,
 author = {Kingma, Durk P and Dhariwal, Prafulla},
 booktitle = {Advances in Neural Information Processing Systems},
 pages = {},
 publisher = {Curran Associates, Inc.},
 title = "{Glow: Generative Flow with Invertible $1\times1$ Convolutions}",
 url = {https://proceedings.neurips.cc/paper_files/paper/2018/file/d139db6a236200b21cc7f752979132d0-Paper.pdf},
 volume = {31},
 year = {2018}
}

@misc{DBLP:journals/corr/XuWCL15,
      title="{Empirical Evaluation of Rectified Activations in Convolutional Network}", 
      author={Bing Xu and Naiyan Wang and Tianqi Chen and Mu Li},
      year={2015},
      eprint={1505.00853},
      archivePrefix={arXiv},
      primaryClass={cs.LG},
      url={https://arxiv.org/abs/1505.00853}, 
}

@misc{papamakarios2017masked,
      title="{Masked Autoregressive Flow for Density Estimation}", 
      author={George Papamakarios and Theo Pavlakou and Iain Murray},
      year={2018},
      eprint={1705.07057},
      archivePrefix={arXiv},
      primaryClass={stat.ML},
      url={https://arxiv.org/abs/1705.07057}, 
}

@inproceedings{DBLP:conf/icml/RezendeM15,
  author       = {Danilo Jimenez Rezende and Shakir Mohamed},
  title        = "{Variational Inference with Normalizing Flows}",
  booktitle = 	 {Proceedings of the 32nd International Conference on Machine Learning},
  pages = 	 {1530--1538},
  year = 	 {2015},
  volume = 	 {37},
  series = 	 {Proceedings of Machine Learning Research},
  address = 	 {Lille, France},
  month = 	 {07--09 Jul},
  publisher =    {PMLR},
}

@article{Alwall:2014hca,
    author = "Alwall, J. and Frederix, R. and Frixione, S. and Hirschi, V. and Maltoni, F. and Mattelaer, O. and Shao, H. -S. and Stelzer, T. and Torrielli, P. and Zaro, M.",
    title = "{The automated computation of tree-level and next-to-leading order differential cross sections, and their matching to parton shower simulations}",
    eprint = "1405.0301",
    archivePrefix = "arXiv",
    primaryClass = "hep-ph",
    reportNumber = "CERN-PH-TH-2014-064, CP3-14-18, LPN14-066, MCNET-14-09, ZU-TH-14-14",
    doi = "10.1007/JHEP07(2014)079",
    journal = "J. High{ }Energy Phys.",
    volume = "07",
    pages = "079",
    year = "2014"
}

@article{Artoisenet:2012st,
    author = "Artoisenet, Pierre and Frederix, Rikkert and Mattelaer, Olivier and Rietkerk, Robbert",
    title = "{Automatic spin-entangled decays of heavy resonances in Monte Carlo simulations}",
    eprint = "1212.3460",
    archivePrefix = "arXiv",
    primaryClass = "hep-ph",
    reportNumber = "NIKHEF-2012-021, CERN-PH-TH-2012-329",
    doi = "10.1007/JHEP03(2013)015",
    journal = "J. High{ }Energy Phys.",
    volume = "03",
    pages = "015",
    year = "2013"
}

@article{Sjostrand:2007gs,
   title="{A brief introduction to PYTHIA 8.1}",
   volume={178},
   ISSN={0010-4655},
   url={http://dx.doi.org/10.1016/j.cpc.2008.01.036},
   DOI={10.1016/j.cpc.2008.01.036},
   number={11},
   journal={Computer Physics Communications},
   publisher={Elsevier BV},
   author={Sjöstrand, Torbjörn and Mrenna, Stephen and Skands, Peter},
   year={2008},
   month={6}}

@article{deFavereau:2013fsa,
    author = "de Favereau, J. and Delaere, C. and Demin, P. and Giammanco, A. and Lema{\^\i}tre, V. and Mertens, A. and Selvaggi, M.",
    collaboration = "DELPHES 3",
    title = "{DELPHES 3, A modular framework for fast simulation of a generic collider experiment}",
    eprint = "1307.6346",
    archivePrefix = "arXiv",
    primaryClass = "hep-ex",
    doi = "10.1007/JHEP02(2014)057",
    journal = "J. High{ }Energy Phys.",
    volume = "02",
    pages = "057",
    year = "2014"
}

@article{Cacciari:2008gp,
    author = "Cacciari, Matteo and Salam, Gavin P. and Soyez, Gregory",
    title = "{The anti-$k_t$ jet clustering algorithm}",
    eprint = "0802.1189",
    archivePrefix = "arXiv",
    primaryClass = "hep-ph",
    reportNumber = "LPTHE-07-03",
    doi = "10.1088/1126-6708/2008/04/063",
    journal = "J. High{ }Energy Phys.",
    volume = "04",
    pages = "063",
    year = "2008"
}

@inproceedings{DBLP:conf/icml/XiongYHZZXZLWL20,
  author       = {Ruibin Xiong and
                  Yunchang Yang and
                  Di He and
                  Kai Zheng and
                  Shuxin Zheng and
                  Chen Xing and
                  Huishuai Zhang and
                  Yanyan Lan and
                  Liwei Wang and
                  Tie{-}Yan Liu},
  title        = "{On Layer Normalization in the Transformer Architecture}",
  booktitle    = {Proceedings of the 37th International Conference on Machine Learning,
                  {ICML} 2020, 13--18 July 2020, Virtual Event},
  series       = {Proceedings of Machine Learning Research},
  volume       = {119},
  pages        = {10524--10533},
  publisher    = {{PMLR}},
  year         = {2020},
  url          = {http://proceedings.mlr.press/v119/xiong20b.html},
  bibsource    = {dblp computer science bibliography, https://dblp.org}
}

@misc{DBLP:journals/corr/abs-2010-11929,
      title="{An Image is Worth $16\times16$ Words: Transformers for Image Recognition at Scale}", 
      author={Alexey Dosovitskiy and Lucas Beyer and Alexander Kolesnikov and Dirk Weissenborn and Xiaohua Zhai and Thomas Unterthiner and Mostafa Dehghani and Matthias Minderer and Georg Heigold and Sylvain Gelly and Jakob Uszkoreit and Neil Houlsby},
      year={2021},
      eprint={2010.11929},
      archivePrefix={arXiv},
      primaryClass={cs.CV},
      url={https://arxiv.org/abs/2010.11929}, 
}

@misc{DBLP:journals/corr/HendrycksG16,
      title="{Gaussian Error Linear Units (GELUs)}", 
      author={Dan Hendrycks and Kevin Gimpel},
      year={2023},
      eprint={1606.08415},
      archivePrefix={arXiv},
      primaryClass={cs.LG},
      url={https://arxiv.org/abs/1606.08415}, 
}

@misc{DBLP:journals/corr/BaKH16,
      title="{Layer Normalization}", 
      author={Jimmy Lei Ba and Jamie Ryan Kiros and Geoffrey E. Hinton},
      year={2016},
      eprint={1607.06450},
      archivePrefix={arXiv},
      primaryClass={stat.ML},
      url={https://arxiv.org/abs/1607.06450}, 
}

@article{DBLP:journals/jmlr/SrivastavaHKSS14,
  author       = {Nitish Srivastava and
                  Geoffrey E. Hinton and
                  Alex Krizhevsky and
                  Ilya Sutskever and
                  Ruslan Salakhutdinov},
  title        = "{Dropout: a simple way to prevent neural networks from overfitting}",
  journal      = {J. Mach. Learn. Res.},
  volume       = {15},
  number       = {1},
  pages        = {1929--1958},
  year         = {2014},
  url          = {https://dl.acm.org/doi/10.5555/2627435.2670313},
  doi          = {10.5555/2627435.2670313},
  bibsource    = {dblp computer science bibliography, https://dblp.org}
}

@inproceedings{NIPS2000_44968aec,
 author = {Dugas, Charles and Bengio, Yoshua and B\'{e}lisle, Fran\c{c}ois and Nadeau, Claude and Garcia, Ren\'{e}},
 booktitle = {Advances in Neural Information Processing Systems},
 editor = {T. Leen and T. Dietterich and V. Tresp},
 pages = {},
 publisher = {MIT Press},
 title = "{Incorporating Second-Order Functional Knowledge for Better Option Pricing}",
 url = {https://proceedings.neurips.cc/paper_files/paper/2000/file/44968aece94f667e4095002d140b5896-Paper.pdf},
 volume = {13},
 year = {2000}
}

@misc{DBLP:journals/corr/HeZRS15,
      title="{Deep Residual Learning for Image Recognition}", 
      author={Kaiming He and Xiangyu Zhang and Shaoqing Ren and Jian Sun},
      year={2015},
      eprint={1512.03385},
      archivePrefix={arXiv},
      primaryClass={cs.CV},
      url={https://arxiv.org/abs/1512.03385}, 
}

@misc{DBLP:journals/corr/HeZR016,
      title="{Identity Mappings in Deep Residual Networks}", 
      author={Kaiming He and Xiangyu Zhang and Shaoqing Ren and Jian Sun},
      year={2016},
      eprint={1603.05027},
      archivePrefix={arXiv},
      primaryClass={cs.CV},
      url={https://arxiv.org/abs/1603.05027}, 
}

@misc{DBLP:journals/corr/abs-1711-05101,
      title="{Decoupled Weight Decay Regularization}", 
      author={Ilya Loshchilov and Frank Hutter},
      year={2019},
      eprint={1711.05101},
      archivePrefix={arXiv},
      primaryClass={cs.LG},
      url={https://arxiv.org/abs/1711.05101}, 
}

@misc{DBLP:journals/corr/LoshchilovH16a,
      title="{SGDR: Stochastic Gradient Descent with Warm Restarts}", 
      author={Ilya Loshchilov and Frank Hutter},
      year={2017},
      eprint={1608.03983},
      archivePrefix={arXiv},
      primaryClass={cs.LG},
      url={https://arxiv.org/abs/1608.03983}, 
}

@misc{DBLP:journals/corr/GoyalDGNWKTJH17,
      title="{Accurate, Large Minibatch SGD: Training ImageNet in 1 Hour}", 
      author={Priya Goyal and Piotr Dollár and Ross Girshick and Pieter Noordhuis and Lukasz Wesolowski and Aapo Kyrola and Andrew Tulloch and Yangqing Jia and Kaiming He},
      year={2018},
      eprint={1706.02677},
      archivePrefix={arXiv},
      primaryClass={cs.CV},
      url={https://arxiv.org/abs/1706.02677}, 
}

@article{liu1989limited,
  title="{On the limited memory BFGS method for large scale optimization}",
  author={Liu, Dong C and Nocedal, Jorge},
  journal={Mathematical programming},
  volume={45},
  number={1},
  pages={503--528},
  year={1989},
  publisher={Springer}
}

@article{wolfe1969convergence,
  title={Convergence conditions for ascent methods},
  author={Wolfe, Philip},
  journal={SIAM review},
  volume={11},
  number={2},
  pages={226--235},
  year={1969},
  publisher={SIAM}
}

@article{wolfe1971convergence,
  title="{Convergence conditions for ascent methods. II: Some corrections}",
  author={Wolfe, Philip},
  journal={SIAM review},
  volume={13},
  number={2},
  pages={185--188},
  year={1971},
  publisher={SIAM}
}

@book{wright1999numerical,
  title="{Numerical Optimization}",
  author={Nocedal, Jorge and Wright, Stephen J.},
  isbn={9780387400655},
  series={Springer Series in Operations Research and Financial Engineering},
  year={2006},
  edition={2nd},
  publisher={Springer},
  address={New York, NY}
}

@article{Evans:1129806,
      author        = "Evans, Lyndon R and Bryant, Philip",
      title         = "{LHC Machine}",
      journal = "{JINST}",
      volume        = "3",
      pages         = "S08001",
      year          = 2008,
      url           = "https://cds.cern.ch/record/1129806",
      doi           = "10.1088/1748-0221/3/08/S08001",
}

@article{ATLAS:2008xda,
    author = "Aad, G. and others",
    collaboration = "ATLAS",
    title = "{The ATLAS Experiment at the CERN Large Hadron Collider}",
    journal = "{JINST}",
    volume = "3",
    pages = "S08003",
    year = 2008,
    doi = "10.1088/1748-0221/3/08/S08003",
}

@article{ATLAS:2015pfy,
    author = "Aad, Georges and others",
    collaboration = "ATLAS",
    title = "{Determination of the top-quark pole mass using $ t\overline{t} $ + 1-jet events collected with the ATLAS experiment in 7 TeV pp collisions}",
    eprint = "1507.01769",
    archivePrefix = "arXiv",
    primaryClass = "hep-ex",
    reportNumber = "CERN-PH-EP-2015-100",
    doi = "10.1007/JHEP10(2015)121",
    journal = "J. High{ }Energy Phys.",
    volume = "10",
    pages = "121",
    year = "2015"
}

@article{CMS:2018quc,
    author = "Sirunyan, Albert M and others",
    collaboration = "CMS",
    title = "{Measurement of the top quark mass with lepton+jets final states using $\mathrm {p}$$\mathrm {p}$ collisions at $\sqrt{s}=13\,\text {TeV} $}",
    eprint = "1805.01428",
    archivePrefix = "arXiv",
    primaryClass = "hep-ex",
    reportNumber = "CMS-TOP-17-007, CERN-EP-2018-063",
    doi = "10.1140/epjc/s10052-018-6332-9",
    journal = "Eur. Phys. J. C",
    volume = "78",
    number = "11",
    pages = "891",
    year = "2018",
    note = "[Erratum: Eur.Phys.J.C 82, 323 (2022)]"
}

@article{Kvita:2018trd,
    author = "Kvita, J.",
    title = "{Study of methods of resolved top quark reconstruction in semileptonic $t\bar{t}$ decay}",
    eprint = "1806.05463",
    archivePrefix = "arXiv",
    primaryClass = "hep-ex",
    doi = "10.1016/j.nima.2018.05.059",
    journal = "Nucl. Instrum. Meth. A",
    volume = "900",
    pages = "84--100",
    year = "2018",
    note = "[Erratum: Nucl.Instrum.Meth.A 1040, 167172 (2022)]"
}

@article{ATLAS:2018fwq,
    author = "Aaboud, Morad and others",
    collaboration = "ATLAS",
    title = "{Measurement of the top quark mass in the $t\bar{t}\rightarrow $ lepton+jets channel from $\sqrt{s}=8$  TeV ATLAS data and combination with previous results}",
    eprint = "1810.01772",
    archivePrefix = "arXiv",
    primaryClass = "hep-ex",
    reportNumber = "CERN-EP-2018-238",
    doi = "10.1140/epjc/s10052-019-6757-9",
    journal = "Eur. Phys. J. C",
    volume = "79",
    number = "4",
    pages = "290",
    year = "2019"
}

@article{ATLAS:2019guf,
    author = "Aad, Georges and others",
    collaboration = "ATLAS",
    title = "{Measurement of the top-quark mass in $t\bar{t}+1$-jet events collected with the ATLAS detector in $pp$ collisions at $\sqrt{s}=8$ TeV}",
    eprint = "1905.02302",
    archivePrefix = "arXiv",
    primaryClass = "hep-ex",
    reportNumber = "CERN-EP-2019-059",
    doi = "10.1007/JHEP11(2019)150",
    journal = "J. High{ }Energy Phys.",
    volume = "11",
    pages = "150",
    year = "2019"
}

@article{ATLAS:2019hxz,
    author = "Aad, Georges and others",
    collaboration = "ATLAS",
    title = "{Measurements of top-quark pair differential and double-differential cross-sections in the $\ell$+jets channel with $pp$ collisions at $\sqrt{s}=13$ TeV using the ATLAS detector}",
    eprint = "1908.07305",
    archivePrefix = "arXiv",
    primaryClass = "hep-ex",
    reportNumber = "CERN-EP-2019-149",
    doi = "10.1140/epjc/s10052-019-7525-6",
    journal = "Eur. Phys. J. C",
    volume = "79",
    number = "12",
    pages = "1028",
    year = "2019",
    note = "[Erratum: Eur.Phys.J.C 80, 1092 (2020)]"
}

@article{Grossi:2020orx,
    author = "Grossi, M. and Novak, J. and Kersevan, B. and Rebuzzi, D.",
    title = "{Comparing traditional and deep-learning techniques of kinematic reconstruction for polarization discrimination in vector boson scattering}",
    eprint = "2008.05316",
    archivePrefix = "arXiv",
    primaryClass = "hep-ph",
    reportNumber = "VBSCAN-PUB-07-20",
    doi = "10.1140/epjc/s10052-020-08713-1",
    journal = "Eur. Phys. J. C",
    volume = "80",
    number = "12",
    pages = "1144",
    year = "2020"
}

@article{CMS:2021vhb,
    author = "Tumasyan, Armen and others",
    collaboration = "CMS",
    title = "{Measurement of differential $t \bar t$ production cross sections in the full kinematic range using lepton+jets events from proton--proton collisions at $\sqrt {s}$ = 13{\,}{\,}TeV}",
    eprint = "2108.02803",
    archivePrefix = "arXiv",
    primaryClass = "hep-ex",
    reportNumber = "CMS-TOP-20-001, CERN-EP-2021-135",
    doi = "10.1103/PhysRevD.104.092013",
    journal = "Phys. Rev. D",
    volume = "104",
    number = "9",
    pages = "092013",
    year = "2021"
}

@article{ATLAS:2022waa,
    author = "Aad, G. and others",
    collaboration = "ATLAS",
    title = "{Evidence for the charge asymmetry in pp {\textrightarrow} $ t\overline{t} $ production at $ \sqrt{s} $ = 13 TeV with the ATLAS detector}",
    eprint = "2208.12095",
    archivePrefix = "arXiv",
    primaryClass = "hep-ex",
    reportNumber = "CERN-EP-2022-166",
    doi = "10.1007/JHEP08(2023)077",
    journal = "J. High{ }Energy Phys.",
    volume = "08",
    pages = "077",
    year = "2023"
}

\end{document}